\documentclass[%
 reprint,
 amsmath,amssymb,twocolumn,prl
]{revtex4-2}

\usepackage{graphicx}
\usepackage{dcolumn}
\usepackage{bm}
\usepackage{makecell}
\usepackage{multirow}
\usepackage{threeparttable}
\usepackage{booktabs}
\newcommand{\Ecell}[3]{%
	\makecell[c]{%
		\ensuremath{#1}\\
		\ensuremath{#2}\\
		\ensuremath{(#3\%)}
	}%
}
\usepackage{siunitx}
\usepackage{amsmath}
\allowdisplaybreaks[4]
\usepackage[sort&compress]{natbib}
\begin{document}

\preprint{APS/123-QED}

\title{The origin of ferroelectricity, polarization and high resistivity in Aurivillius CaBi$_{2}$B$_{2}$O$_{9}$ (B = Ta, Nb) }

\author{Fengzhang Tang}
\author{Shi Wei}
\author{Qi Hu}
\author{Jie Xing}
\author{Jianguo Zhu}
\author{Zhi Tan}
\thanks{Corresponding authors: tanzhi0838@scu.edu.cn and qcscu@scu.edu.cn}

\author{Qiang Chen}
\thanks{Corresponding authors: tanzhi0838@scu.edu.cn and qcscu@scu.edu.cn}

\affiliation{College of Materials Science and Engineering, Sichuan University, Chengdu 610065, China}

\date{\today}

\begin{abstract}
Aurivillius layered oxides are important candidates for high-temperature ferroelectric and piezoelectric application. In this work, we combine group theoretic analysis with first-principles calculations to systematically investigate the origin of ferroelectric phase transition, polarization, piezoelectric response, and intrinsic electrical insulation of the two-layer Aurivillius ferroelectrics CaBi$_{2}$B$_{2}$O$_{9}$ (B = Ta, Nb). The results show that the \textit{A2$_1$am} ferroelectric phase arises from the cooperative condensation of a polar mode and nonpolar oxygen octahedral rotation/tilting modes, whose the $\Gamma_5^-$X$_2^+$X$_3^-$ trilinear coupling substantially lowers the total energy and deepens the ferroelectric potential well. The spontaneous polarization and anisotropic piezoelectric response are governed primarily by the cooperative displacements of the Bi$_{2}$O$_{2}$ layers and Ta/NbO$_{6}$ octahedra, with Bi ions providing an indispensable contribution to both responses. More importantly, the polar distortion can be traced to the relative in-plane displacement between adjacent the Bi$_{2}$O$_{2}$ layer and the perovskite-like block. Because this displacement is intrinsic to the alternating Bi$_{2}$O$_{2}$/perovskite-block stacking topology and is independent of the number of perovskite layers, we identify interlayer sliding as a general, layer-number-independent structural mechanism for ferroelectricity in Aurivillius oxides. Band structure and effective-mass analyses further reveal that the wide O 2p-Ta/Nb d charge-transfer gap and weakly dispersive band edges jointly give rise to their high intrinsic resistivity. Our findings establish a unified microscopic picture linking structural distortions, ferroelectric polarization, piezoelectric response, and electronic insulation in CaBi$_{2}$B$_{2}$O$_{9}$ (B = Ta, Nb), and provide theoretical guidance for designing layered ferroelectric oxides with high Curie temperatures and robust insulating behavior.  

\end{abstract}

\maketitle

\section{introduction}
Ferroelectric materials are polar insulators that possess a spontaneous electric polarization whose direction can be reversibly switched by an external electric field. This switchable polarization makes ferroelectrics attractive for nonvolatile memories, capacitors, pyroelectric detectors, electro-optic modulators, and other field-tunable electronic devices\cite{1}. Beyond their switchable polarization, ferroelectrics constitute the most important class of piezoelectric materials. In polycrystalline ferroelectric ceramics, electrical poling aligns ferroelectric domains and establishes a macroscopic non-centrosymmetric state, thereby generating a net piezoelectric response. The resulting electromechanical coupling allows the interconversion of mechanical and electrical energy and forms the basis for numerous technologies, including actuators, resonators, ultrasonic transducers, and energy-harvesting systems\cite{2}. The growing need for sensing and monitoring in harsh environments has stimulated extensive research into high-temperature piezoelectric materials. To ensure reliable operation at elevated temperatures, these materials must possess a high Curie temperature, low dielectric loss, and high electrical resistivity. Such characteristics are crucial for mitigating leakage-current-induced depolarization and maintaining stable electromechanical performance under extreme conditions\cite{3}.

Aurivillius layered oxides constitute an important family of ferroelectric materials that consist of perovskite-like slabs interleaved with Bi$_{2}$O$_{2}$ layers along the crystallographic direction. Their general formula can be written as Bi$_{2m}$A$_{n-m}$B$_{n}$O$_{3(m+n)}$, where n denotes the number of octahedral layers in the perovskite block. The A site is usually occupied by large cations, such as Ca$^{2+}$, Sr$^{2+}$, Ba$^{2+}$, Pb$^{2+}$, Bi$^{3+}$. The B site is occupied by sixfold-coordinated cations, such as Fe$^{3+}$, Cr$^{3+}$, Ti$^{4+}$, Ta$^{5+}$, or Nb$^{5+}$\cite{4,5}. The intrinsic layered structure introduces strong crystallographic anisotropy and ferroelectric behaviors distinct from those of conventional ABO$_{3}$ perovskites. Among Aurivillius oxides, CaBi$_{2}$Ta$_{2}$O$_{9}$ (CBTO) and CaBi$_{2}$Nb$_{2}$O$_{9}$ (CBNO) are representative two-layer ferroelectric compounds with exceptionally high Curie temperatures of approximately 1196 K\cite{6}and 1213 K\cite{7}, respectively. Owing to their outstanding thermal stability and excellent electrical insulation, these materials have attracted considerable attention as promising candidates for high-temperature piezoelectric applications. Nevertheless, their exceptionally high Curie temperatures are rather intriguing when compared with those of conventional perovskite tantalates and niobates, despite the fact that they contain the same fundamental structural units, namely TaO$_{6}$ and NbO$_{6}$ octahedra. For example, KNbO$_{3}$ undergoes a ferroelectric phase transition near 673 K, whereas KTaO$_{3}$ remains quantum paraelectric down to 0 K. According to the conventional understanding of perovskite ferroelectrics, ferroelectricity is primarily driven by hybridization between transition-metal d$^{0}$ states and the oxygen 2p states\cite{8}, and the spontaneous polarization originates mainly from the off-center displacement of B-site cations. Furthermore, both ferroelectricity and cation off-centering are generally suppressed under external pressure\cite{9,10}. From this perspective, the presence of the relatively small Ca$^{2+}$ cation at the A-site in CBTO and CBNO is expected to impose stronger chemical pressure than the larger K$^{+}$ cation in KNbO$_{3}$ and KTaO$_{3}$, which would ordinarily be unfavorable for the stabilization of ferroelectricity. These apparent contradictions raise several fundamental questions. What is the microscopic origin of ferroelectricity in CBTO and CBNO? Which atomic displacements contribute most significantly to the spontaneous polarization? More importantly, what mechanism is responsible for their extraordinarily high Curie temperatures? The answers to these questions are likely rooted in the unique layered architecture of Aurivillius oxides. Nevertheless, the manner in which this structural feature governs ferroelectricity remains poorly understood. 

Although the ferroelectric, piezoelectric, and high-temperature insulating properties of CBTO and CBNO-based ceramics have been improved by chemical substitution\cite{11,12,13}, cooperative site modification\cite{14}, and texturing\cite{15,16,17}, these experimental studies mainly focus on performance optimization and leave the underlying microscopic mechanisms unclear. First-principles studies on related Aurivillius oxides have revealed that their ferroelectricity often arises from the coupling between polar distortions and nonpolar oxygen-octahedral rotation/tilting modes, rather than from a single polar soft mode\cite{18,19,20,21,22}. However, a unified microscopic picture connecting the ferroelectric phase transition, high Curie temperature, spontaneous polarization, piezoelectric response, and intrinsic high resistivity remains lacking for CBTO and CBNO. In particular, despite their similar crystal structures and comparably high Curie temperatures, CBTO and CBNO exhibit substantially different piezoelectric responses and electrical resistivities, the microscopic origins of which remain unresolved. A deeper understanding of these issues is essential not only for elucidating the fundamental ferroelectric mechanisms in Aurivillius oxides, but also for guiding the development of high-performance piezoelectric materials capable of operating under extreme conditions.

In this study, taking the \textit{I4/mmm} phase as the high-symmetry reference paraelectric structure and the \textit{A2$_1$am} phase as the ferroelectric ground state, we identify the primary structural distortion modes by group-theoretical mode decomposition and lattice-dynamical analysis, and further examine the coupling between the polar mode and nonpolar oxygen-octahedral rotation/tilting modes using a multimode Landau free-energy expansion. The spontaneous polarization is calculated using the Berry-phase method, and its microscopic origin is analyzed in terms of Born effective charges. The piezoelectric strain coefficients and their microscopic contributions are investigated using the finite-stress method. Finally, the origin of the high resistivity in these two compounds is discussed based on their band structures and carrier effective masses. This study aims to establish a unified microscopic physical picture linking structural distortions, polarization, piezoelectricity, and electronic insulation in CaBi$_{2}$B$_{2}$O$_{9}$ (B = Ta, Nb), and to provide theoretical guidance for designing layered ferroelectric materials with high Curie temperatures and robust insulating behavior.

\section{methods}
All first-principles calculations are performed within density functional theory using the Vienna Ab initio Simulation Package (VASP)\cite{23}. The projection augmented wave (PAW) method is used to describe the interaction between the ionic cores and valence electrons\cite{24,25,26}, and the Perdew-Burke-Ernzerhof functional (PBE) is used within the generalized gradient approximation to treat the exchange-correlation energy\cite{27}. The cutoff energy of the plane wave is set to 520 eV. The Brillouin zone is sampled by Monkhorst-Pack k-point meshes with a reciprocal-space resolution of 2$\pi$ × 0.03 \AA$^{-1}$. Both lattice parameters and internal atomic coordinates are fully relaxed until the total energy converged to less than 10$^{-5}$ eV and the residual force on each atom is below 10$^{-2}$ eV/\AA. Phonon dispersion relations are calculated using the finite-displacement method implemented in the Phonopy package\cite{28}, with 3×3×1 supercells. The high-symmetry paths in the Brillouin zone are generated using SeeK-path\cite{29}. The band structures and densities of states are calculated using the PBE, PBEsol\cite{30} and SCAN\cite{31} exchange-correlation functionals. For the group theoretic analysis, symmetry-adapted lattice distortion modes are decomposed using the AMPLIMODES program available on the Bilbao Crystallographic Server\cite{32}, and the Landau-type energy expansion is constructed with the aid of INVARIANTS\cite{33}. The ferroelectric polarization is calculated using the Berry-phase method\cite{34,35}, and Born effective charges are used to analyze the microscopic origin of the polarization. The piezoelectric tensor is evaluated by a finite-difference approach\cite{36}. The chemical bonding analysis is carried out by utilizing LOBSTER within the framework of crystal orbital Hamiltonian population (COHP)\cite{37,38}.
\begin{figure*}[btp]
	\includegraphics[width=\linewidth]{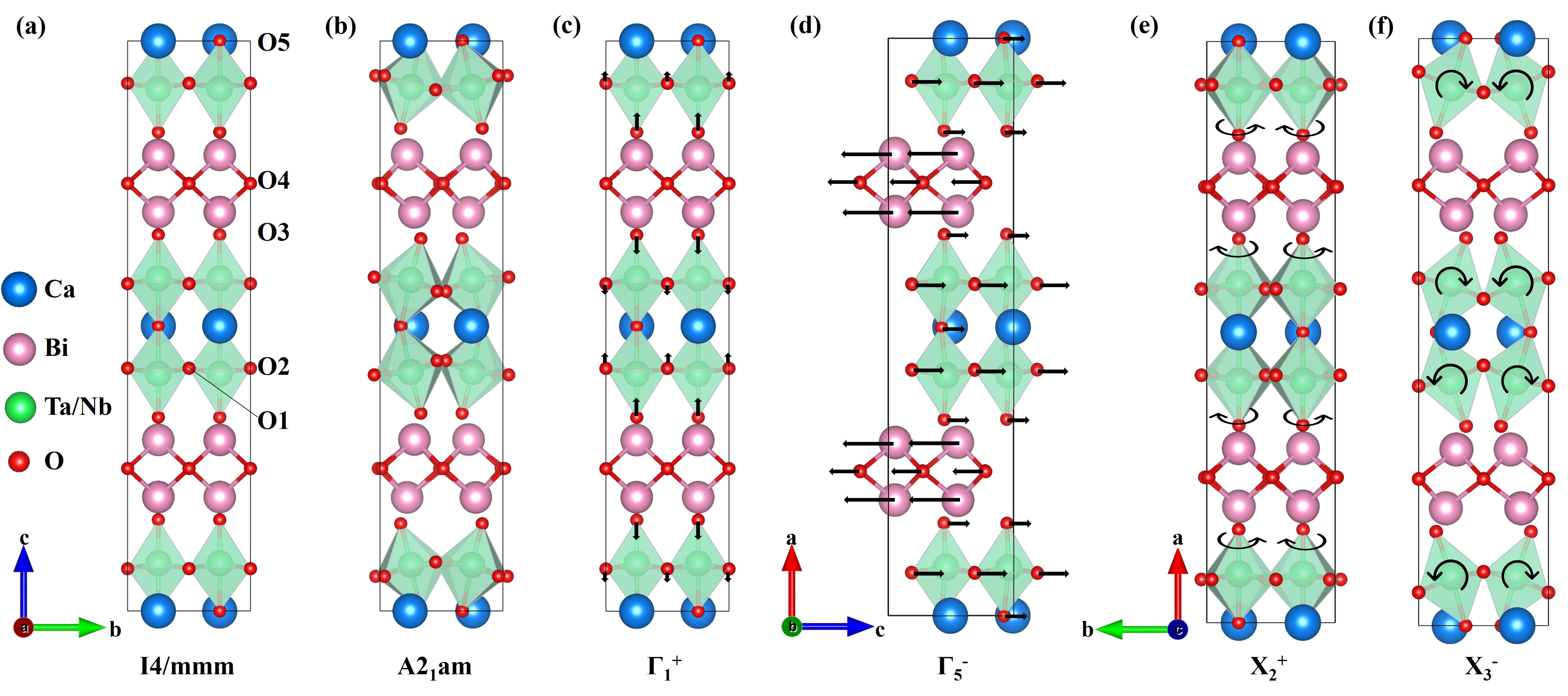}
	\caption{\label{fig1}  Crystal structures and symmetry-adapted distortion modes of CaBi$_{2}$B$_{2}$O$_{9}$ (B = Ta, Nb). (a) Tetragonal \textit{I4/mmm} parent structure and (b) polar orthorhombic \textit{A2$_1$am} structure. The transformation from the \textit{I4/mmm} parent phase to the \textit{A2$_1$am} phase is described in terms of four symmetry-adapted lattice modes: (c) the fully symmetric $\Gamma_1^+$ relaxation mode, (d) the polar $\Gamma_5^-$ mode, (e) the octahedral rotation mode X$_2^+$, and (f) the octahedral tilt mode X$_3^-$. The arrows denote the characteristic atomic displacements of each distortion. Blue, pink, green, and red spheres represent Ca, Bi, B-site cations (B=Ta, Nb), and O atoms, respectively.}
\end{figure*} 

\begin{table*}[t]
	\centering
	
	\caption{
		Symmetry-adapted lattice distortion mode decomposition of the
		${A2_{1}am}$ phase of
		$\mathrm{CaBi_2Ta_2O_9}$ and $\mathrm{CaBi_2Nb_2O_9}$.
		The structure is obtained by fully relaxing the atomic positions
		and lattice parameters using the GGA-PBE functional.
		The mode amplitudes are reported in the so-called
		``parentcell-normalized'' values of AMPLIMODES in the
		Bilbao Crystallographic Server.
		``Amplitude$^{*}$'' represents the length of the global minimum
		point of the fitted Landau multi-mode energy landscape.
	}
	
	\label{tab:mode_decomposition}
	
	\small
	\renewcommand{\arraystretch}{1.30}
	\setlength{\tabcolsep}{6pt}
	
	\begin{tabular*}{\textwidth}{
			@{\extracolsep{\fill}}
			l
			l
			c
			c
			c
			c
			c
			@{}
		}
		
		\specialrule{0.8pt}{0pt}{1.2pt}
		\specialrule{0.4pt}{0pt}{2pt}
		
		\multicolumn{2}{c}{Materials}
		&
		$\Gamma_{1}^{+}$
		&
		$\Gamma_{5}^{-}$
		&
		$X_{2}^{+}$
		&
		$X_{3}^{-}$
		&
		Overall
		\\
		
		\midrule
		
		\multirow[c]{3}{*}{$\mathrm{CaBi_2Ta_2O_9}$}
		&
		Amplitude (\AA)
		&
		0.2734
		&
		0.9367
		&
		0.6306
		&
		1.0287
		&
		1.5517
		\\
		
		&
		Amplitude$^{*}$ (\AA)
		&
		0.2510
		&
		0.9009
		&
		0.6058
		&
		0.9673
		&
		\\
		
		&
		Percentage (\%)
		&
		3.10\%
		&
		36.43\%
		&
		16.51\%
		&
		43.95\%
		&
		\\
		
		\midrule
		
		\multirow[c]{3}{*}{$\mathrm{CaBi_2Nb_2O_9}$}
		&
		Amplitude (\AA)
		&
		0.2813
		&
		0.9729
		&
		0.6704
		&
		1.0518
		&
		1.6067
		\\
		
		&
		Amplitude$^{*}$ (\AA)
		&
		0.2385
		&
		0.9720
		&
		0.6325
		&
		0.9590
		&
		\\
		
		&
		Percentage (\%)
		&
		3.07\%
		&
		36.67\%
		&
		17.41\%
		&
		42.86\%
		&
		\\
		
		\specialrule{0.4pt}{2pt}{1.2pt}
		\specialrule{0.8pt}{0pt}{0pt}
		
	\end{tabular*}
	
\end{table*}

\section{results}
\subsection{Multimode-coupling mechanism for the ferroelectric phase transition}
\begin{figure*}[btp]
	\includegraphics[width=\linewidth]{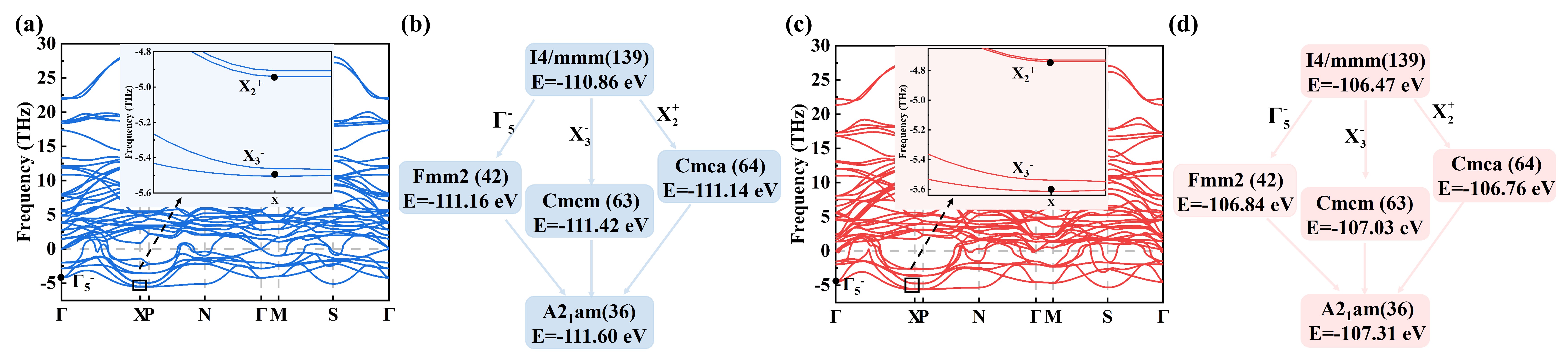}
	\caption{\label{fig2} Phonon dispersion curves and structural energetics of CaBi$_{2}$B$_{2}$O$_{9}$ (B = Ta, Nb). Calculated phonon dispersions of the tetragonal \textit{I4/mmm} parent phase for (a) CaBi$_{2}$Ta$_{2}$O$_{9}$ and (c) CaBi$_{2}$Nb$_{2}$O$_{9}$. The insets show enlarged views of the unstable modes near the X point, where the X$_2^+$ and X$_3^-$ instabilities are identified. 
		Panels (b) and (d) show the symmetry-related metastable phases generated by freezing in the unstable modes of CaBi$_{2}$Ta$_{2}$O$_{9}$ and CaBi$_{2}$Nb$_{2}$O$_{9}$ , respectively. The arrows denote the structural pathways driven by the polar $\Gamma_5^-$ distortion, the octahedral rotation X$_2^+$, and the octahedral tilt X$_3^-$. The calculated total energies listed in the boxes indicate that the combined distortion leads to the lowest-energy polar \textit{A2$_1$am} phase.}
\end{figure*}

The fully relaxed paraelectric phase \textit{I4/mmm} (No.139) and ferroelectric phase \textit{A2$_1$am} (No.36) of CaBi$_{2}$Ta$_{2}$O$_{9}$ (CBTO) and CaBi$_{2}$Nb$_{2}$O$_{9}$ (CBNO) are shown in Fig. 1. Some studies denote this ferroelectric phase as \textit{Cmc2$_1$}. The \textit{A2$_1$am} and \textit{Cmc2$_1$} descriptions correspond to the same space group No.36, and represent the same orthorhombic polar structure, differing only in the choice of crystallographic setting or unit cell. In this work, we consistently use \textit{A2$_1$am} to denote the ferroelectric phase. To clarify the structural evolution from the high-symmetry paraelectric phase to the low-symmetry ferroelectric phase, we perform a symmetry-mode decomposition using the Bilbao Crystallographic Server. The results show that this phase transition involves four dominant symmetry-adapted lattice distortion modes [Fig~\ref{fig2}(c-d) and Table I]: (1) $\Gamma_1^+$ mode, a fully symmetric relaxation mode that does not by itself lower the symmetry of the parent phase and is usually induced as a secondary mode by other primary distortions; (2) $\Gamma_5^-$ mode, a polar displacement mode that directly breaks inversion symmetry and gives rise to a polar \textit{Fmm2} structure (No.42); (3) X$_2^+$ mode, correspond to in-phase oxygen-octahedral rotation, which leads to the \textit{Cmca} (No.64) phase; (4) X$_3^-$ mode, correspond to anti-phase oxygen-octahedral tilting distortions, which generates the \textit{Cmcm} (No.63) phase.

The mode-decomposition results show that the contributions of the X$_3^-$, $\Gamma_5^-$, X$_2^+$, and $\Gamma_1^+$ modes in CBTO are 43.95 \%, 36.43 \%, 16.51 \%, and 3.10 \% for total distortion, respectively. The corresponding values in CBNO are 42.86 \%, 36.67 \%, 17.41 \%, and 3.07 \% [Table I]. Thus, the two compounds exhibit nearly identical distortion compositions. Among these modes, X$_3^-$ mode gives the largest contribution, indicating that oxygen octahedral tilting is the dominant structural degree of freedom stabilizing the \textit{A2$_1$am} phase. This is consistent with the small geometrical tolerance factor (\textit{t} = 0.950) of the perovskite-like slab, since octahedral tilting is a common mechanism for releasing lattice mismatch and lowering the energy in structures with (\textit{t} \textless 1) \cite{39,40}. Meanwhile, the sizable $\Gamma_5^-$ component indicates that polar displacement is also an essential ingredient in the formation of the low-symmetry ferroelectric phase. Notably, the smaller $\Gamma_5^-$ distortion amplitude in CBTO compared with that in CBNO should be consistent with its lower spontaneous polarization observed in experiments. Compared with CBTO, CBNO has a slightly larger total distortion amplitude, suggesting that Nb substitution for Ta enhances the structural deviation from the high-symmetry parent phase. To elucidate the microscopic origin of ferroelectricity, the phonon dispersions of the high-symmetry \textit{I4/mmm} phase are investigated. Pronounced imaginary-frequency soft modes are observed in both compounds, revealing a strong dynamical instability of the tetragonal parent phase and suggesting that the ferroelectric transition is driven by soft-mode condensation [Fig. ~\ref{fig2}(a) and 2(c)]. Among the unstable phonon modes, the X$_3^-$ mode has the largest imaginary frequency, followed by the X$_2^+$ and $\Gamma_5^-$ modes. The presence of these strongly unstable modes indicates that the ferroelectric state of CBTO and CBNO does not arise from the condensation of a single distortion mode. Instead, it is likely stabilized through the coupling and cooperative interaction of the X$_3^-$, X$_2^+$, and $\Gamma_5^-$ modes.

\begin{table*}[t]
	\centering
	
	\caption{
		Fitted coefficients for the Landau energy expansions of the
		$A2_{1}am$ phase [Eq.~(1)] of
		$\mathrm{CaBi_2Ta_2O_9}$ and $\mathrm{CaBi_2Nb_2O_9}$
		at fixed cubic lattice parameters.
		The coefficients are fitted using the units of energy in meV/f.u.\
		and mode amplitudes in \AA.
	}
	
	\label{tab:landau_coefficients}
	
	\vspace{2pt}
	
	\small
	\renewcommand{\arraystretch}{1.35}
	\setlength{\tabcolsep}{6pt}
	
	\resizebox{\textwidth}{!}{%
		\begin{tabular}{@{}cccccc@{}}
			
			\specialrule{0.8pt}{0pt}{1.2pt}
			\specialrule{0.4pt}{0pt}{2pt}
			
			Phase
			&
			\multicolumn{5}{c}{Fitted coefficients}
			\\
			
			\midrule
			
			\multirow{7}{*}{
				\makecell[c]{
					$\mathrm{CaBi_2Ta_2O_9}$\\[2pt]
					$A2_{1}am$
				}
			}
			&
			$a_{\Gamma_1^{+}}=17.1$
			&
			$A_{\Gamma_1^{+}}=3025.0$
			&
			$b_{\Gamma_1^{+}}=590.8$
			&
			$B_{\Gamma_1^{+}}=72.2$
			&
			$D_{\Gamma_1^{+}\Gamma_5^{-}}=566.3$
			\\
			
			&
			&
			$A_{\Gamma_5^{-}}=-376.7$
			&
			$\lambda_{\Gamma_1^{+}\Gamma_5^{-}}=-747.7$
			&
			$B_{\Gamma_5^{-}}=221.8$
			&
			$D_{\Gamma_1^{+}X_2^{+}}=531.3$
			\\
			
			&
			&
			$A_{X_2^{+}}=-573.3$
			&
			$\lambda_{\Gamma_1^{+}X_2^{+}}=-663.5$
			&
			$B_{X_2^{+}}=442.1$
			&
			$D_{\Gamma_1^{+}X_3^{-}}=301.1$
			\\
			
			&
			&
			$A_{X_3^{-}}=-785.7$
			&
			$\lambda_{\Gamma_1^{+}X_3^{-}}=-1382.5$
			&
			$B_{X_3^{-}}=432.3$
			&
			$D_{\Gamma_5^{-}X_2^{+}}=313.0$
			\\
			
			&
			&
			&
			$C_{\Gamma_5^{-}X_2^{+}X_3^{-}}=-764.3$
			&
			&
			$D_{\Gamma_5^{-}X_3^{-}}=313.6$
			\\
			
			&
			&
			&
			&
			&
			$D_{X_2^{+}X_3^{-}}=703.5$
			\\
			
			&
			&
			&
			&
			&
			$\beta_{\Gamma_1^{+}\Gamma_5^{-}X_2^{+}X_3^{-}}=107.1$
			\\
			
			\midrule
			
			\multirow{7}{*}{
				\makecell[c]{
					$\mathrm{CaBi_2Nb_2O_9}$\\[2pt]
					$A2_{1}am$
				}
			}
			&
			$a_{\Gamma_1^{+}}=61.2$
			&
			$A_{\Gamma_1^{+}}=3257.5$
			&
			$b_{\Gamma_1^{+}}=766.7$
			&
			$B_{\Gamma_1^{+}}=-517.9$
			&
			$D_{\Gamma_1^{+}\Gamma_5^{-}}=439.2$
			\\
			
			&
			&
			$A_{\Gamma_5^{-}}=-460.4$
			&
			$\lambda_{\Gamma_1^{+}\Gamma_5^{-}}=-706.7$
			&
			$B_{\Gamma_5^{-}}=262.4$
			&
			$D_{\Gamma_1^{+}X_2^{+}}=952.6$
			\\
			
			&
			&
			$A_{X_2^{+}}=-527.1$
			&
			$\lambda_{\Gamma_1^{+}X_2^{+}}=-769.7$
			&
			$B_{X_2^{+}}=415.9$
			&
			$D_{\Gamma_1^{+}X_3^{-}}=332.4$
			\\
			
			&
			&
			$A_{X_3^{-}}=-756.6$
			&
			$\lambda_{\Gamma_1^{+}X_3^{-}}=-1411.1$
			&
			$B_{X_3^{-}}=445.5$
			&
			$D_{\Gamma_5^{-}X_2^{+}}=360.5$
			\\
			
			&
			&
			&
			$C_{\Gamma_5^{-}X_2^{+}X_3^{-}}=-994.7$
			&
			&
			$D_{\Gamma_5^{-}X_3^{-}}=277.4$
			\\
			
			&
			&
			&
			&
			&
			$D_{X_2^{+}X_3^{-}}=729.3$
			\\
			
			&
			&
			&
			&
			&
			$\beta_{\Gamma_1^{+}\Gamma_5^{-}X_2^{+}X_3^{-}}=258.3$
			\\
			
			\specialrule{0.4pt}{2pt}{1.2pt}
			\specialrule{0.8pt}{0pt}{0pt}
			
		\end{tabular}%
	}
	
\end{table*}

To quantitatively describe the interactions among these modes, we use $\Gamma_1^+$, $\Gamma_5^-$, X$_2^+$, and X$_3^-$ as order parameters and write the Landau-type energy expansion up to fourth order as \cite{41}:
\begin{align}
	E_{\mathrm{A}2_{1}\mathrm{am}}
	={}& E_{\mathrm{I}4/\mathrm{mmm}}
	+\mathrm{a}_{\Gamma_{1}^{+}}Q_{\Gamma_{1}^{+}}
	+A_{\Gamma_{1}^{+}}Q_{\Gamma_{1}^{+}}^{2}
	+\mathrm{b}_{\Gamma_{1}^{+}}Q_{\Gamma_{1}^{+}}^{3}
	\notag\\
	&+B_{\Gamma_{1}^{+}}Q_{\Gamma_{1}^{+}}^{4}
	+A_{\Gamma_{5}^{-}}Q_{\Gamma_{5}^{-}}^{2}
	+A_{X_{2}^{+}}Q_{X_{2}^{+}}^{2}
	\notag\\
	&+A_{X_{3}^{-}}Q_{X_{3}^{-}}^{2}
	+\lambda_{\Gamma_{1}^{+}\Gamma_{5}^{-}}
	Q_{\Gamma_{1}^{+}}Q_{\Gamma_{5}^{-}}^{2}
	\notag\\
	&+\lambda_{\Gamma_{1}^{+}X_{2}^{+}}
	Q_{\Gamma_{1}^{+}}Q_{X_{2}^{+}}^{2}
	+\lambda_{\Gamma_{1}^{+}X_{3}^{-}}
	Q_{\Gamma_{1}^{+}}Q_{X_{3}^{-}}^{2}
	\notag\\
	&+C_{\Gamma_{5}^{-}X_{2}^{+}X_{3}^{-}}
	Q_{\Gamma_{5}^{-}}Q_{X_{2}^{+}}Q_{X_{3}^{-}}
	\notag\\
	&+B_{\Gamma_{5}^{-}}Q_{\Gamma_{5}^{-}}^{4}
	+B_{X_{2}^{+}}Q_{X_{2}^{+}}^{4}
	+B_{X_{3}^{-}}Q_{X_{3}^{-}}^{4}
	\notag\\
	&+D_{\Gamma_{1}^{+}\Gamma_{5}^{-}}
	Q_{\Gamma_{1}^{+}}^{2}Q_{\Gamma_{5}^{-}}^{2}
	+D_{\Gamma_{1}^{+}X_{2}^{+}}
	Q_{\Gamma_{1}^{+}}^{2}Q_{X_{2}^{+}}^{2}
	\notag\\
	&+D_{\Gamma_{1}^{+}X_{3}^{-}}
	Q_{\Gamma_{1}^{+}}^{2}Q_{X_{3}^{-}}^{2}
	+D_{\Gamma_{5}^{-}X_{2}^{+}}
	Q_{\Gamma_{5}^{-}}^{2}Q_{X_{2}^{+}}^{2}
	\notag\\
	&+D_{\Gamma_{5}^{-}X_{3}^{-}}
	Q_{\Gamma_{5}^{-}}^{2}Q_{X_{3}^{-}}^{2}
	+D_{X_{2}^{+}X_{3}^{-}}
	Q_{X_{2}^{+}}^{2}Q_{X_{3}^{-}}^{2}
	\notag\\
	&+\beta_{\Gamma_{1}^{+}\Gamma_{5}^{-}X_{2}^{+}X_{3}^{-}}
	Q_{\Gamma_{1}^{+}}
	Q_{\Gamma_{5}^{-}}
	Q_{X_{2}^{+}}
	Q_{X_{3}^{-}}
	\label{eqn:eps}
\end{align}
where ${{E}_{I4/\text{mmm}}}$ is the energy of the tetragonal reference structure, ${{Q}_{\Gamma _{1}^{+}}}$, ${{Q}_{\Gamma _{5}^{-}}}$, ${{Q}_{X_{2}^{+}}}$ and ${{Q}_{X_{3}^{-}}}$are the mode amplitudes of the $\Gamma_1^+$, $\Gamma_5^-$, X$_2^+$, and X$_3^-$, respectively, and the a, A, b, B, $\lambda$, C, D parameters are the coefficients in the energy expansion, with the subscripts specifying the corresponding modes. The coefficients are obtained by introducing structural distortions with different mode amplitudes at the fixed lattice parameters of the parent phase and performing a global least-squares fit to the first-principles total energies [Table II]. The fitting quality (R$^2$ \textgreater 0.99) indicates that this energy expansion provides a reliable description of the relevant potential-energy surface. The fitted coefficients show that, in both compounds, the quadratic coefficients of the $\Gamma_5^-$, X$_2^+$, and X$_3^-$ modes are negative, while their quartic coefficients are positive. Such a coefficient combination gives rise to double-well energy profiles, indicating that all three modes are unstable with respect to the high-symmetry parent structure and spontaneously condense to finite distortion amplitudes. Among them, the X$_3^-$ mode has the most negative quadratic coefficient, further confirming that anti-phase oxygen octahedral tilting represents the strongest structural instability. In contrast, the quadratic coefficient of the $\Gamma_1^+$ mode is positive, indicating that this distortion is not intrinsically unstable in the parent phase. Nevertheless, it can be induced through coupling with other primary distortions and contribute to lowering the total energy via linear–quadratic coupling terms in the free-energy expansion. Importantly, the linear–quadratic coupling coefficients associated with the $\Gamma_1^+$ mode are found to be significantly negative in both CBTO and CBNO. Therefore, although the amplitude of the $\Gamma_1^+$ distortion is relatively small, its contribution to the stability of the ferroelectric phase cannot be neglected. More importantly, the Landau expansion contains a pronounced trilinear coupling term $\Gamma_5^-$X$_2^+$X$_3^-$. The corresponding coefficients are -764.3 and -994.7 for CBTO and CBNO, respectively, indicating that the total energy is further lowered when the polar displacement, octahedral rotation, and octahedral tilting condense cooperatively with the appropriate relative phases. Meanwhile, most biquadratic coupling coefficients are positive, suggesting a certain competition in amplitude among different modes. Therefore, the ferroelectric phase is not stabilized solely by an intrinsic polar soft mode. Its stabilization results from the combined effects of trilinear-coupling-driven energy lowering, biquadratic-coupling-induced amplitude competition, and quartic anharmonic terms.

\begin{table*}[t]
	\centering
	
	\caption{The energy contributions (meV/f.u.) of each relevant term
		in the Landau free-energy expression to the total energy reduction.}
	\label{tab:energy_contributions}
	
	\small
	\renewcommand{\arraystretch}{1.15}
	
	\begin{tabular*}{\textwidth}
		{@{\extracolsep{\fill}}llcccccc@{}}
		
		\specialrule{0.8pt}{0pt}{1.2pt}
		\specialrule{0.4pt}{0pt}{2pt}
		
		\multicolumn{8}{c}
		{Contributions for lowering energy (meV/f.u.)}
		\\
		
		\midrule
		
		\multirow[c]{4}{*}[-30pt]{
			$\mathrm{CaBi_2Ta_2O_9}$
		}
		&
		\makecell[c]{Single\\mode}
		&
		\Ecell{\Gamma_1^{+}}{233.9}{-30.9}
		&
		\Ecell{\Gamma_5^{-}}{-159.8}{21.1}
		&
		\Ecell{X_2^{+}}{-158.1}{20.9}
		&
		\Ecell{X_3^{-}}{-347.3}{45.9}
		&
		&
		\\
		
		\cmidrule(l){2-8}
		
		&
		\makecell[c]{Biquadratic\\coupling\\term}
		&
		\Ecell{\Gamma_1^{+}\Gamma_5^{-}}{-142.2}{18.8}
		&
		\Ecell{\Gamma_1^{+}X_2^{+}}{-56.3}{7.5}
		&
		\Ecell{\Gamma_1^{+}X_3^{-}}{-376.2}{49.7}
		&
		\Ecell{\Gamma_5^{-}X_2^{+}}{109.2}{-14.4}
		&
		\Ecell{\Gamma_5^{-}X_3^{-}}{291.2}{-38.5}
		&
		\Ecell{X_2^{+}X_3^{-}}{296.0}{-39.1}
		\\
		
		\cmidrule(l){2-8}
		
		&
		\makecell[c]{Trilinear\\coupling\\term}
		&
		\Ecell{\Gamma_5^{-}X_2^{+}X_3^{-}}{-464.4}{61.4}
		&
		&
		&
		&
		&
		\\
		
		\cmidrule(l){2-8}
		
		&
		\makecell[c]{Four-linear\\coupling\\term}
		&
		\Ecell{\Gamma_1^{+}\Gamma_5^{-}X_2^{+}X_3^{-}}{17.8}{-2.3}
		&
		&
		&
		&
		&
		\\
		
		\midrule
		
		\multirow[c]{4}{*}[-30pt]{
			$\mathrm{CaBi_2Nb_2O_9}$
		}
		&
		\makecell[c]{Single\\mode}
		&
		\Ecell{\Gamma_1^{+}}{254.4}{-30.1}
		&
		\Ecell{\Gamma_5^{-}}{-200.7}{23.7}
		&
		\Ecell{X_2^{+}}{-152.9}{18.1}
		&
		\Ecell{X_3^{-}}{-291.8}{34.5}
		&
		&
		\\
		
		\cmidrule(l){2-8}
		
		&
		\makecell[c]{Biquadratic\\coupling\\term}
		&
		\Ecell{\Gamma_1^{+}\Gamma_5^{-}}{-155.3}{18.4}
		&
		\Ecell{\Gamma_1^{+}X_2^{+}}{-63.4}{7.5}
		&
		\Ecell{\Gamma_1^{+}X_3^{-}}{-410.0}{48.5}
		&
		\Ecell{\Gamma_5^{-}X_2^{+}}{153.4}{-18.1}
		&
		\Ecell{\Gamma_5^{-}X_3^{-}}{290.5}{-34.3}
		&
		\Ecell{X_2^{+}X_3^{-}}{362.6}{-42.9}
		\\
		
		\cmidrule(l){2-8}
		
		&
		\makecell[c]{Trilinear\\coupling\\term}
		&
		\Ecell{\Gamma_5^{-}X_2^{+}X_3^{-}}{-682.4}{80.7}
		&
		&
		&
		&
		&
		\\
		
		\cmidrule(l){2-8}
		
		&
		\makecell[c]{Four-linear\\coupling\\term}
		&
		\Ecell{\Gamma_1^{+}\Gamma_5^{-}X_2^{+}X_3^{-}}{49.9}{-5.9}
		&
		&
		&
		&
		&
		\\
		
		\specialrule{0.4pt}{2pt}{1.2pt}
		\specialrule{0.8pt}{0pt}{0pt}
		
	\end{tabular*}
\end{table*}

\begin{figure*}[btp]
	\includegraphics[width=\linewidth]{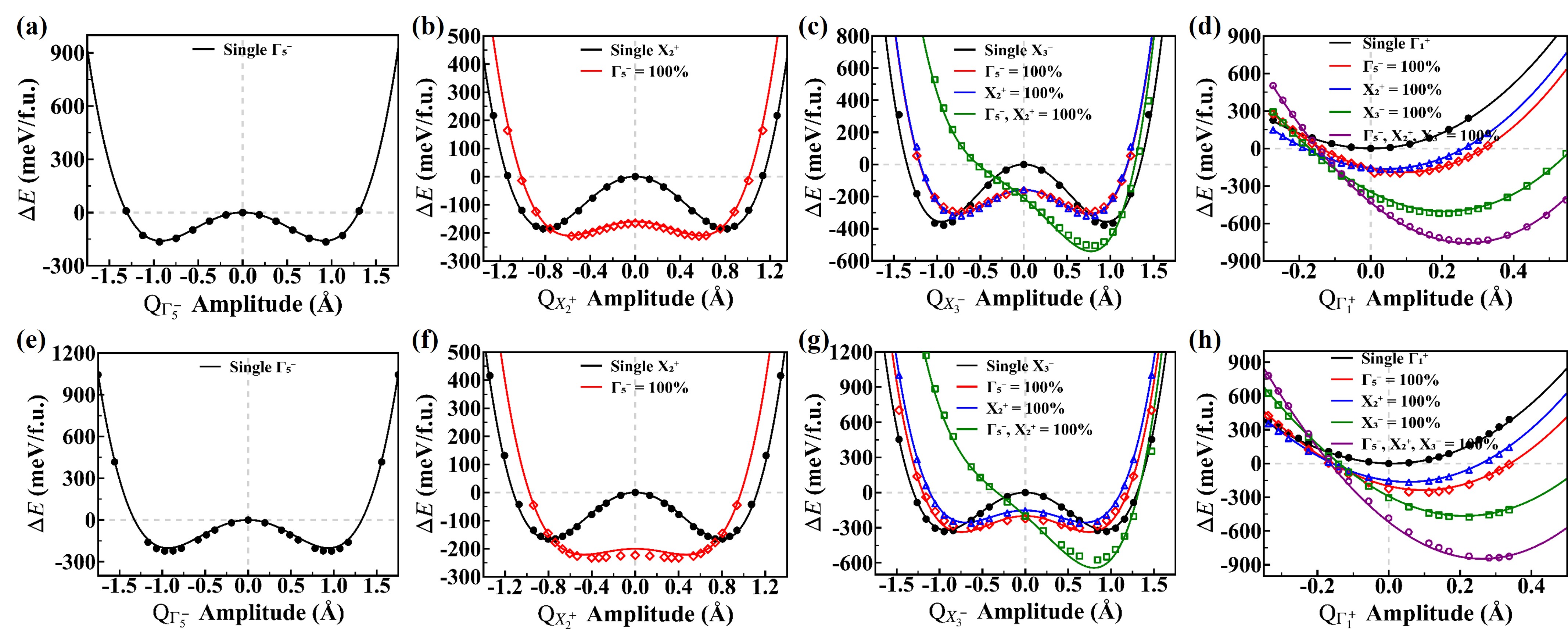}
	\caption{\label{fig3} Frozen-phonon energy profiles showing the energy change $\textit {$\Delta$E}$ as a function of mode amplitude for (a-d) CaBi$_{2}$Ta$_{2}$O$_{9}$ and (e-h) CaBi$_{2}$Nb$_{2}$O$_{9}$, referenced to the tetragonal parent phase. The energy landscapes are shown for the $\Gamma_5^-$, X$_2^+$, X$_3^-$, and $\Gamma_1^+$ modes, respectively. Black curves correspond to single-mode distortions, whereas colored curves show the corresponding energy profiles when other symmetry modes are frozen in at their full amplitudes, as specified in the legends. Symbols denote first-principles data points, and solid lines are polynomial fits.}
\end{figure*} 

The energy curves and the energy decompositions provide a consistent physical picture [Fig. ~\ref{fig3} and Table III]. When the $\Gamma_5^-$, X$_2^+$, or X$_3^-$ mode is activated individually, the system exhibits a double-well energy profile, indicating that each of these modes can independently lower the energy of the parent phase. Among them, the X$_3^-$ mode produces the deepest double well, consistent with the strongest structural instability. The single mode contributions of the $\Gamma_5^-$, X$_2^+$, and X$_3^-$ modes to the energy lowering in CBTO are -159.8, -158.1, and -347.3 meV/f.u., respectively. The corresponding values in CBNO are -200.7, -152.9, and -291.8 meV/f.u. These results indicate that both polar displacements and antiferrodistortive octahedral distortions play important roles in stabilizing the ferroelectric phase. A comparison between the two compounds further reveals distinct mode-dependent stabilization behaviors. The X$_3^-$ mode contributes a more negative energy in CBTO than in CBNO, whereas the $\Gamma_5^-$ mode in CBNO shows a stronger stabilizing effect than that of CBTO. In contrast, the energy contribution from the X$_2^+$ mode is comparable in both systems. These differences suggest that the relative balance between polar and antiferrodistortive instabilities is intrinsically different in the two compounds, which may be closely related to their ferroelectric transition temperatures and piezoelectric response. However, although the $\Gamma_5^-$, X$_2^+$, and X$_3^-$ modes are individually unstable in the paraelectric phase, the all the biquadratic coupling terms contribute positively to the total energy, indicating a strong competition between pairs of distortion modes. Furthermore, when the $\Gamma_5^-$, X$_2^+$, and X$_3^-$ modes coexist, the energy profile becomes asymmetric and develops a deeper minimum along a specific distortion direction, directly reflecting the reconstruction of the multidimensional potential energy surface by the trilinear coupling. The energy decomposition further shows that the $\Gamma_5^-$X$_2^+$X$_3^-$ trilinear term contributes -464.4 meV/f.u. in CBTO, accounting for 61.4\% of the total stabilization energy. In CBNO, this term contributes -682.4 meV/f.u., corresponding to 80.7\% of the total stabilization energy. In contrast, the negative energy contributions arising from the individual unstable modes are largely compensated by the positive contributions associated with the biquadratic coupling terms. Consequently, the trilinear coupling cannot be regarded as a mere perturbative correction to the single-mode instabilities. Instead, it acts as the dominant interaction governing the stabilization of the ferroelectric phase by profoundly reshaping the multidimensional energy landscape. These results demonstrate that the ferroelectric ground states of CBTO and CBNO originate primarily from the cooperative trilinear interaction among the $\Gamma_5^-$, X$_2^+$, and X$_3^-$ distortions. Moreover, the energy lowering associated with the trilinear coupling is substantially larger than that arising from any individual distortion mode. This result indicates that the negative trilinear coupling energetically favors the cooperative condensation of the $\Gamma_5^-$, X$_2^+$, and X$_3^-$ modes, suppressing the stabilization of intermediate states associated with individual distortions. In addition, the $\Gamma_1^+$ mode should not be neglected despite its relatively small distortion amplitude. The isolated $\Gamma_1^+$ mode exhibits a single-well energy profile with its minimum located near zero amplitude, indicating that it is stable in the high-symmetry phase. However, when the three primary unstable modes are present together, its equilibrium position shifts to a finite amplitude, indicating that $\Gamma_1^+$ participates in the stabilization of the ferroelectric phase as an induced fully symmetric relaxation response. Furthermore, the linear-quadratic coupling terms associated with the $\Gamma_1^+$ mode provide a substantial contribution to the total energy lowering. Remarkably, the combined energy gain arising from these coupling terms is comparable to that contributed by the $\Gamma_5^-$X$_2^+$X$_3^-$ trilinear coupling. This result indicates that, although $\Gamma_1^+$ is not a primary instability, it plays an essential role in the stabilization of the ferroelectric ground state through its strong coupling with the primary order parameters.

We now can address the microscopic origin of ferroelectricity in CBTO and CBNO. The parent paraelectric phase of both compounds exhibits multiple structural instabilities. Although the polar $\Gamma_5^-$ mode is unstable, it is not the dominant instability. Instead, the X$_3^-$ mode, corresponding to anti-phase octahedral tilting, exhibits the largest imaginary frequency and therefore represents the primary structural instability of the high-symmetry phase. Furthermore, the strong $\Gamma_5^-$X$_2^+$X$_3^-$ trilinear coupling provides substantial energy lowering, which energetically favors the cooperative condensation of the three distortion modes. As a result, the ferroelectric ground state does not originate from an isolated polar soft mode, but rather from the synergistic interaction among polar displacements, in-phase octahedral rotations, and anti-phase octahedral tilts. Based on the above conclusions, the high Curie temperatures of CBTO and CBNO, as well as their stability over a wide temperature range, can be understood within the framework of the potential-energy landscape. The Curie temperature of a ferroelectric is closely related to the depth of the double-well potential of the ferroelectric phase relative to the paraelectric phase, which reflects the strength of the underlying structural instabilities and their coupling interactions. As shown in Table III, the ferroelectric phases of CBTO and CBNO are lower in energy than the parent phases by 756.2 and 845.7 meV/f.u., respectively. These energy differences are remarkably large compared with those of conventional perovskite ferroelectrics, which typically exhibit stabilization energies on the order of only ~10 meV per atom \cite{9,42}. Such exceptionally deep energy wells provide a natural explanation for the extraordinarily high Curie temperatures and the robust ferroelectric stability observed in both compounds. A large thermal energy is required to destabilize the coupled distortion network and recover the high-symmetry paraelectric structure. Consequently, the ferroelectric order remains robust over a broad temperature range, giving rise to the exceptional thermal stability characteristic of these Aurivillius oxides. The decomposition of the stabilization energy into individual Landau terms provides further insight into the origin of the exceptionally high Curie temperatures in CBTO and CBNO, as well as the differences between the two compounds. Notably, the polar $\Gamma_5^-$ distortion alone accounts for only about 20\% of the total energy lowering. The remaining stabilization energy originates primarily from the X$_2^+$ and X$_3^-$ octahedral distortions and, more importantly, from their coupling interactions with the other distortion modes. This result indicates that the X$_2^+$ and X$_3^-$ octahedral distortions play a crucial role in stabilizing the ferroelectric phase, leading to a mechanism that differs fundamentally from that of conventional perovskite ferroelectrics. 

In classical perovskite ABO$_{3}$ ferroelectrics, the stabilization energy primarily originates from long-range ordered off-center displacements of the B-site cations driven by d$^{0}$ hybridization. Consequently, the magnitude of the polar distortion is generally correlated with the depth of the ferroelectric potential well and thus with the Curie temperature. For example, BaTiO$_{3}$ exhibits a Curie temperature of approximately 400 K, whereas substitution of Ba by smaller A-site cations, such as Sr or Ca, progressively reduces the lattice volume, weakens the polar displacement, and lowers the Curie temperature. In contrast, the high Curie temperatures of CBTO and CBNO do not arise primarily from a strong polar instability. Instead, they originate from the cooperative stabilization provided by the X$_2^+$ and X$_3^-$ octahedral distortions and their coupling interactions, which generate an exceptionally deep multidimensional energy landscape. Consequently, the Curie temperature is governed not only by the polar $\Gamma_5^-$ mode but also strongly by the strength of the octahedral rotation and tilting distortions. The X$_2^+$ and X$_3^-$ distortions are closely associated with the geometric mismatch within the perovskite-like framework, which can be qualitatively described by the tolerance factor \textit{t}. In general, a smaller tolerance factor corresponds to a larger mismatch between the A-site cation and the BO$_{6}$ framework, thereby enhancing the amplitudes of octahedral rotations and tilts. As a result, the stabilization energy associated with these distortions increases, leading to a deeper ferroelectric energy well and a higher Curie temperature. Consequently, the variation of Curie temperature in Aurivillius ferroelectrics differs fundamentally from that in conventional perovskite ferroelectrics. For example, CBTO exhibits a Curie temperature of approximately 1196 K, whereas SrBi$_{2}$Ta$_{2}$O$_{9}$ and BaBi$_{2}$Ta$_{2}$O$_{9}$ possess significantly lower Curie temperatures of about 573 K and 333 K, respectively\cite{43}. This trend is consistent with the progressive increase in the A-site ionic radius from Ca$^{2+}$ to Sr$^{2+}$ to Ba$^{2+}$, which weakens the octahedral rotation and tilting instabilities and consequently reduces the stability of the ferroelectric phase. These findings have important implications for the design of high-temperature Aurivillius piezoelectrics. In particular, the incorporation of large-radius A-site cations should be carefully considered, as such substitutions may weaken the octahedral rotation and tilting instabilities that are essential for stabilizing the ferroelectric phase and maintaining a high Curie temperature. 

This mechanism also provides a natural explanation for why CBTO and CBNO exhibit nearly identical Curie temperatures, in sharp contrast to the markedly different behaviors of KTaO$_{3}$ and KNbO$_{3}$. Because both compounds contain the same A-site cation, Ca$^{2+}$, they possess very similar octahedral rotation and tilting amplitudes, resulting in comparable stabilization energies and consequently nearly identical Curie temperatures. Under this mechanism, the Curie temperature is governed primarily by the stability of the octahedral framework and its coupling interactions, while the contribution of the polar mode becomes comparatively less dominant. Nevertheless, the polar $\Gamma_5^-$ mode remains intrinsically unstable in both compounds and therefore plays an indispensable role in the formation of the ferroelectric ground state. More importantly, the microscopic origin of the polar distortion in Aurivillius oxides differs fundamentally from that in conventional perovskite ferroelectrics. As will be shown in the following section, the layered crystal structure of the Aurivillius phase provides a distinct route for generating spontaneous polarization, leading to a ferroelectric mechanism that is markedly different from the classical B-site off-centering picture.

\begin{figure*}[btp]
	\includegraphics[width=\linewidth]{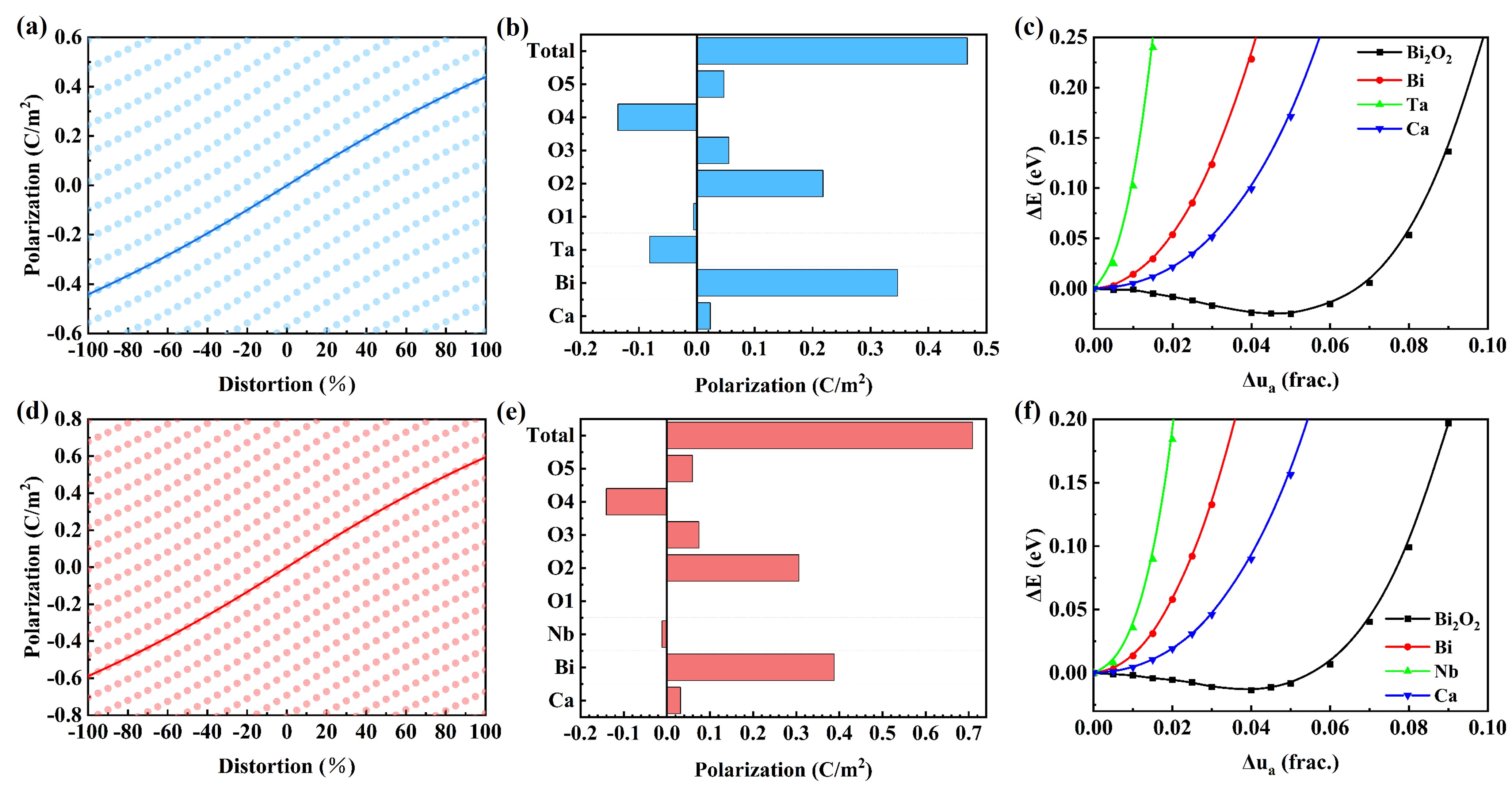}
	\caption{\label{fig4} Berry-phase polarization as a function of structural distortion along the switching path from the antiferroelectric state to the centrosymmetric paraelectric state and then to the ferroelectric state for (a) CaBi$_{2}$Ta$_{2}$O$_{9}$ and (d) CaBi$_{2}$Nb$_{2}$O$_{9}$. Panels (b) and (e) show the decomposition of the total polarization into contributions from the inequivalent atomic sublattices for CaBi$_{2}$Ta$_{2}$O$_{9}$ and CaBi$_{2}$Nb$_{2}$O$_{9}$, respectively. Panels (c) and (f) show the frozen-displacement energy profiles for selected atoms and structural units along the polar a direction for CaBi$_{2}$Ta$_{2}$O$_{9}$ and CaBi$_{2}$Nb$_{2}$O$_{9}$, respectively, where $\textit {$\Delta$E}$ is referenced to the centrosymmetric \textit{I4/mmm} phase.}
\end{figure*} 

\subsection{Polarization properties}
We first calculate the spontaneous polarization using the standard Berry-phase method based on the modern theory of polarization [Fig.~\ref{fig4}(a) and (d)]. The high-symmetry \textit{I4/mmm} paraelectric phase is used as the reference structure. An adiabatic distortion path is then constructed between the \textit{I4/mmm} phase and the polar \textit{A2$_1$am} phase. Along this path, 0 \% corresponds to the high-symmetry parent phase, while the +100 \% and -100 \% structures correspond to two ferroelectric states with opposite polarization directions. In a periodic crystal, the polarization is multivalued due to the polarization quantum. Therefore, the Berry-phase polarization must be tracked on the same physical branch along a continuous adiabatic path. The spontaneous polarization is then obtained from the polarization difference between the two opposite ferroelectric states:
\begin{equation}
	{{P}_{s}}=\frac{1}{2}\left| P(+100\%)-P(-100\%) \right|\
	\label{eqn:DOs}
\end{equation}

The results show that the polarization of CBTO and CBNO changes approximately linearly with the distortion amplitude, indicating that the polar distortion path is continuous and also showing that no obvious jump occurs between different polarization branches. The spontaneous polarizations are 0.44 C/m$^{2}$ for CBTO and 0.59 C/m$^{2}$ for CBNO, indicating substantial ferroelectric polarization in both compounds. Notably, the polarization of CBNO is larger than that of CBTO. This is consistent with the distortion-mode decomposition analysis, which shows that CBNO possesses a larger amplitude of the polar $\Gamma_5^-$ mode. The enhanced $\Gamma_5^-$ distortion therefore contributes directly to the larger spontaneous polarization observed in CBNO.

\begin{table*}[t]
	\centering
	
	\caption{
		Layer-resolved displacement of
		$\mathrm{CaBi_2Ta_2O_9}$ and $\mathrm{CaBi_2Nb_2O_9}$.
		$N_i$ is the number of atoms at each crystallographic site in the
		computational cell.
		$\Delta u_a$ denotes the total atomic displacement along the polar
		$a$ direction from the centrosymmetric $I4/mmm$ reference phase to
		the ferroelectric $A2_{1}am$ phase.
		$\Delta U_{\mathrm{L}}$ is the displacement of the center of the
		corresponding structural layer, and $\delta u_a$ is the internal
		displacement relative to the layer center.
		$Z_{aa}^{*}$ is the Born effective charge component calculated for
		the high-symmetry reference phase.
	}
	
	\label{tab:layer_resolved_displacement}
	
	\small
	\renewcommand{\arraystretch}{1.22}
	\setlength{\tabcolsep}{5pt}
	
	\begin{tabular*}{\textwidth}{
			@{\extracolsep{\fill}}
			l
			l
			c
			c
			c
			c
			c
			c
			@{}
		}
		
		\specialrule{0.8pt}{0pt}{1.2pt}
		\specialrule{0.4pt}{0pt}{2pt}
		
		Materials
		&
		\makecell[c]{Structural\\layers}
		&
		\makecell[c]{Atomic\\sites}
		&
		$N_i$
		&
		\makecell[c]{$\Delta u_a$\\(frac.)}
		&
		\makecell[c]{$\Delta U_{\mathrm{L}}$\\(frac.)}
		&
		\makecell[c]{$\delta u_a$\\(frac.)}
		&
		$Z_{aa}^{*}$
		\\
		
		\midrule
		
		\multirow[c]{8}{*}{$\mathrm{CaBi_2Ta_2O_9}$}
		&
		$\mathrm{Bi_2O_2}$
		&
		$\mathrm{Bi}$
		&
		8
		&
		0.080
		&
		0.065
		&
		0.015
		&
		4.714
		\\
		
		&
		$\mathrm{Bi_2O_2}$
		&
		$\mathrm{O4}$
		&
		8
		&
		0.050
		&
		0.065
		&
		-0.015
		&
		-2.959
		\\
		
		&
		$\mathrm{CaTa_2O_7}$
		&
		$\mathrm{Ca}$
		&
		4
		&
		0.020
		&
		-0.026
		&
		0.046
		&
		2.493
		\\
		
		&
		$\mathrm{CaTa_2O_7}$
		&
		$\mathrm{Ta}$
		&
		8
		&
		-0.014
		&
		-0.026
		&
		0.012
		&
		6.515
		\\
		
		&
		$\mathrm{CaTa_2O_7}$
		&
		$\mathrm{O1}$
		&
		8
		&
		0.002
		&
		-0.026
		&
		0.028
		&
		-3.210
		\\
		
		&
		$\mathrm{CaTa_2O_7}$
		&
		$\mathrm{O2}$
		&
		8
		&
		-0.074
		&
		-0.026
		&
		-0.048
		&
		-3.189
		\\
		
		&
		$\mathrm{CaTa_2O_7}$
		&
		$\mathrm{O3}$
		&
		8
		&
		-0.028
		&
		-0.026
		&
		-0.002
		&
		-2.139
		\\
		
		&
		$\mathrm{CaTa_2O_7}$
		&
		$\mathrm{O5}$
		&
		4
		&
		-0.052
		&
		-0.026
		&
		-0.026
		&
		-1.955
		\\
		
		\midrule
		
		\multirow[c]{8}{*}{$\mathrm{CaBi_2Nb_2O_9}$}
		&
		$\mathrm{Bi_2O_2}$
		&
		$\mathrm{Bi}$
		&
		8
		&
		0.082
		&
		0.065
		&
		0.017
		&
		5.114
		\\
		
		&
		$\mathrm{Bi_2O_2}$
		&
		$\mathrm{O4}$
		&
		8
		&
		0.049
		&
		0.065
		&
		-0.017
		&
		-3.106
		\\
		
		&
		$\mathrm{CaNb_2O_7}$
		&
		$\mathrm{Ca}$
		&
		4
		&
		0.026
		&
		-0.026
		&
		0.052
		&
		2.600
		\\
		
		&
		$\mathrm{CaNb_2O_7}$
		&
		$\mathrm{Nb}$
		&
		8
		&
		-0.001
		&
		-0.026
		&
		0.025
		&
		8.529
		\\
		
		&
		$\mathrm{CaNb_2O_7}$
		&
		$\mathrm{O1}$
		&
		8
		&
		0.000
		&
		-0.026
		&
		0.026
		&
		-4.204
		\\
		
		&
		$\mathrm{CaNb_2O_7}$
		&
		$\mathrm{O2}$
		&
		8
		&
		-0.079
		&
		-0.026
		&
		-0.052
		&
		-4.204
		\\
		
		&
		$\mathrm{CaNb_2O_7}$
		&
		$\mathrm{O3}$
		&
		8
		&
		-0.034
		&
		-0.026
		&
		-0.008
		&
		-2.341
		\\
		
		&
		$\mathrm{CaNb_2O_7}$
		&
		$\mathrm{O5}$
		&
		4
		&
		-0.058
		&
		-0.026
		&
		-0.032
		&
		-2.178
		\\
		
		\specialrule{0.4pt}{2pt}{1.2pt}
		\specialrule{0.8pt}{0pt}{0pt}
		
	\end{tabular*}
	
\end{table*}

\begin{table*}[t]
	\centering
	
	\caption{
		Layer-center decomposition of the Born-effective-charge linear
		polarization for $\mathrm{CaBi_2Ta_2O_9}$ and
		$\mathrm{CaBi_2Nb_2O_9}$.
		$P_{\mathrm{total}}=P_{\mathrm{shift}}+P_{\mathrm{intra}}$:
		$P_{\mathrm{shift}}$ represents the polarization contribution from
		the rigid displacement of each structural layer, while
		$P_{\mathrm{intra}}$ represents the contribution from internal polar
		distortion within each layer.
		The polarization contributions are evaluated using the Born effective
		charges of the $I4/mmm$ reference phase.
		The site-resolved polarization contribution is estimated as
		$P_i=(16.0218/\Omega)\sum_i N_i Z_{aa}^{*}\delta u_a$,
		where $\Omega$ is the unit-cell volume in \AA$^{3}$.
		The fractional displacements are converted into Cartesian displacements
		using $a=5.540$~\AA\ for $\mathrm{CaBi_2Ta_2O_9}$ and
		$a=5.533$~\AA\ for $\mathrm{CaBi_2Nb_2O_9}$.
		All polarization values are given in $\mathrm{C\,m^{-2}}$.
	}
	
	\label{tab:layer_center_polarization}
	
	\small
	\renewcommand{\arraystretch}{1.25}
	\setlength{\tabcolsep}{6pt}
	
	\begin{tabular*}{\textwidth}{
			@{\extracolsep{\fill}}
			c
			c
			c
			c
			c
			@{}
		}
		
		\specialrule{0.8pt}{0pt}{1.2pt}
		\specialrule{0.4pt}{0pt}{2pt}
		
		Materials
		&
		Structural layers
		&
		$P_{\mathrm{shift}}$
		&
		$P_{\mathrm{intra}}$
		&
		$P_{\mathrm{total}}$
		\\
		
		\midrule
		
		\multirow[c]{3}{*}{$\mathrm{CaBi_2Ta_2O_9}$}
		&
		$\mathrm{Bi_2O_2}$
		&
		0.105
		&
		0.106
		&
		0.212
		\\
		
		&
		$\mathrm{CaTa_2O_7}$
		&
		0.042
		&
		0.211
		&
		0.253
		\\
		
		&
		Total
		&
		0.146
		&
		0.317
		&
		0.465
		\\
		
		\midrule
		
		\multirow[c]{3}{*}{$\mathrm{CaBi_2Nb_2O_9}$}
		&
		$\mathrm{Bi_2O_2}$
		&
		0.121
		&
		0.130
		&
		0.248
		\\
		
		&
		$\mathrm{CaNb_2O_7}$
		&
		0.049
		&
		0.412
		&
		0.464
		\\
		
		&
		Total
		&
		0.170
		&
		0.542
		&
		0.712
		\\
		
		\specialrule{0.4pt}{2pt}{1.2pt}
		\specialrule{0.8pt}{0pt}{0pt}
		
	\end{tabular*}
	
\end{table*}

To further clarify the microscopic origin of the polarization, we decompose the polarization using the Born effective charges of the \textit{I4/mmm} phase. Within the linear-response approximation, the polarization can be written as:
\begin{equation}
	{{P}_{\text{i}}}=\frac{e}{\Omega }\sum\limits_{\kappa ,\alpha }{Z_{\kappa \text{i}\alpha }^{*}}\Delta {{\mu }_{\kappa \alpha }}\
	\label{eqn:Fano}
\end{equation}
where $\Omega$ is the unit-cell volume, $Z_{\kappa \alpha \beta }^{*}$ is the Born effective charge tensor of the k$_{th}$ atom, $\Delta {{\mu }_{\kappa \alpha }}$ is the displacement of this atom from the \textit{I4/mmm} reference structure to the \textit{A2$_1$am} ferroelectric structure. This decomposition only describes the linear response around the high-symmetry phase. Within the present decomposition framework, Bi is found to contribute the largest fraction of the spontaneous polarization, as shown in [Fig.~\ref{fig4}(b) and 4(e)]. This result highlights the crucial role of Bi displacements in the ferroelectric behavior of Aurivillius oxides and reveals a microscopic polarization mechanism that differs fundamentally from conventional story of perovskite ferroelectrics. Returning to [Fig.~\ref{fig1}], the polar $\Gamma_5^-$ distortion can be viewed as a relative sliding between Bi$_{2}$O$_{2}$ layer and perovskite block. As summarized in Table IV, the decomposed atomic displacements $\delta$u$_{a}$ exhibit a highly consistent displacement direction within each structural unit. Specifically, the Bi and O4 atoms in the Bi$_{2}$O$_{2}$ layer undergo positive displacements, whereas the Ta (or Nb), O2, O3, and O5 atoms in the perovskite block exhibit negative displacements, consistent with the characteristic displacement pattern of the $\Gamma_5^-$ mode. It is important to note that neither the Bi$_{2}$O$_{2}$ layer nor the perovskite block is individually charge neutral. The total Born effective charges of the Bi$_{2}$O$_{2}$ layer and the perovskite block are +3.51 and -3.51, respectively, in CBTO. Consequently, their relative sliding gives rise to a substantial spontaneous polarization and constitutes the dominant polarization mechanism in Aurivillius ferroelectrics. However, this picture alone cannot fully account for the difference in spontaneous polarization between CBTO and CBNO. For example, although both Bi and O4 belong to the Bi$_{2}$O$_{2}$ layer, their displacement amplitudes differ significantly (0.08 and 0.05 frac., respectively), indicating the existence of internal relative displacements within the Bi$_{2}$O$_{2}$ layer itself. Similar internal distortions are also present within the perovskite block, where the relative displacements among the B-site cations and oxygen atoms resemble the off-centering mechanism commonly observed in conventional perovskite ferroelectrics. Therefore, the spontaneous polarization in Aurivillius oxides should be understood as arising from two cooperative components: the relative sliding between the charged Bi$_{2}$O$_{2}$ layers and perovskite blocks, and the internal polar distortions within each structural unit. The latter is expected to play a key role in determining the polarization difference between CBTO and CBNO.

To gain deeper insight into the polarization mechanism of layered Aurivillius structures, we introduce a layer-center decomposition scheme to quantitatively analyze the microscopic origin of the polarization [Table IV and V]. In contrast to a simple decomposition into atomic-sublattice contributions, the total displacement of each atom is written as the sum of the rigid displacement of the corresponding structural layer and the internal displacement relative to the layer center, i.e., $\Delta$u$_{i}$=$\Delta$U$_{L}$+$\delta$u$_{i}$. Accordingly, the linear polarization can be separated into a layer-shift contribution P$_{shift}$, and an intralayer distortion contribution P$_{intra}$. Here, P$_{shift}$ originates from the rigid displacements of the Bi$_{2}$O$_{2}$ layer and the CaB$_{2}$O$_{7}$ perovskite-like block, and can be equivalently viewed as the contribution associated with the relative sliding between the two charged structural units. In contrast, Pintra describes the relative off-centering between cations and oxygen sublattices within each structural block.

We first note that the rigid layer displacement, $\Delta$U$_{L}$, which represents the relative sliding between the Bi$_{2}$O$_{2}$ layer and the perovskite block, has nearly the same magnitude in both CBTO and CBNO. This result indicates that the interlayer-sliding contribution to the spontaneous polarization is essentially identical in the two compounds. The layer-shift contribution, P$_{shift}$, is calculated to be 0.146 C/m$^{2}$ and 0.170 C/m$^{2}$ for CBTO and CBNO, respectively. These values are substantially smaller than the corresponding total spontaneous polarizations, indicating that the relative sliding between the Bi$_{2}$O$_{2}$ layer and the perovskite block alone cannot account for the observed polarization. Instead, the intralayer distortions provide the dominant contribution to the spontaneous polarization in both compounds. Despite the relatively short Bi-O4 bond length (typically ~2.3 \AA), indicating strong Bi-O bonding, the Bi$_{2}$O$_{2}$ layer does not behave as a perfectly rigid unit. Instead, it still exhibits an appreciable internal polar distortion, with an intralayer displacement, $\delta$u$_{a}$, of approximately 0.015 and 0.017 frac. for CBTO and CBNO, respectively. Owing to the nearly identical intralayer displacements, the intralayer polarization contributions from the Bi$_{2}$O$_{2}$ layers are also very similar, amounting to 0.106 and 0.130 C/m$^{2}$ for CBTO and CBNO, respectively. Thus, the polarization difference originates predominantly from the perovskite-like block. The intralayer displacement $\delta$u$_{a}$ of Nb is more than twice that of Ta. This enhanced off-center displacement can be attributed to the stronger tendency of Nb through its 4d configuration, which has long been recognized as the microscopic origin of ferroelectricity in Nb-based perovskite oxides. The intralayer polarization contribution P$_{shift}$ from the perovskite block is calculated to be 0.211 and 0.412 C/m$^{2}$ for CBTO and CBNO, respectively. This pronounced difference accounts for most of the disparity in the spontaneous polarization between the two compounds, indicating that the distinct polar distortions within the perovskite block are primarily responsible for their different polarization behaviors. It should also be emphasized that the polarization associated with the Bi$_{2}$O$_{2}$ layer, including both the interlayer-sliding P$_{shift}$ and intralayer-distortion P$_{intra}$ contributions, remains substantial. The total contribution from the Bi$_{2}$O$_{2}$ layer accounts for approximately 45.5\% and 34.8\% of the spontaneous polarization in CBTO and CBNO, respectively. Notably, the Bi ions themselves provide the dominant positive contribution, corresponding to approximately 75\% and 55\% of the total spontaneous polarization in CBTO and CBNO, respectively. This contribution is partially compensated by the negative contribution from the O4 atoms within the Bi$_{2}$O$_{2}$ layer, resulting in the smaller net polarization associated with the Bi$_{2}$O$_{2}$ layer. These results indicate that the Bi$_{2}$O$_{2}$ layer plays a much more significant role in the polarization of Aurivillius ferroelectrics than is commonly assumed in the experimental literature, where its contribution has often been underestimated because the Bi$_{2}$O$_{2}$ layer was treated as a rigid structural unit or adopted as the displacement reference. 

Based on the above analysis, the microscopic origin of spontaneous polarization in Aurivillius ferroelectrics can be classified into three distinct components: (i) the relative sliding between the two oppositely charged structural units, (ii) the intralayer polar distortion within the Bi$_{2}$O$_{2}$ layer, and (iii) the intralayer polar distortion within the perovskite block. Among these, the third component is analogous to the conventional off-centering mechanism found in ABO$_{3}$ perovskite ferroelectrics, whereas the first two are unique characteristics of the layered Aurivillius structure. The above analysis naturally raises a fundamental question: what is the microscopic origin of emergence polarization in Aurivillius oxides? First, the intrinsic polar instability of the perovskite-like block alone is unlikely to be the primary driving force. This conclusion is supported by the fact that Ta-based ABO$_{3}$ perovskites, such as KTaO$_{3}$, NaTaO$_{3}$, and AgTaO$_{3}$, generally do not exhibit pronounced ferroelectric distortions under ambient conditions and retain centrosymmetric crystal structures. Therefore, the TaO$_{6}$ framework itself cannot account for the emergence of polarization in CBTO. Instead, the polarization originates from the unique layered architecture of the Aurivillius structure, in which the Bi-containing Bi$_{2}$O$_{2}$ layers should play a crucial role. From the perspective of atomic displacements, the rigid relative sliding between the Bi$_{2}$O$_{2}$ layer and the perovskite block is much larger than the Bi$_{2}$O$_{2}$ intralayer distortions. This suggests that the emergence of the polar structure is closely associated with the sliding distortion. 

We calculate the frozen-displacement energy profiles for selected atomic sublattices and structural units along the polar direction [Fig.~\ref{fig4}(c) and 4(f)]. Starting from the high-symmetry \textit{I4/mmm} reference structure, a selected atom or structural unit is displaced along the polar direction while all the remaining degrees of freedom are fixed, which compare the local energetic response of different polar displacement channels. For both CBTO and CBNO, the energy increases rapidly when the B-site Ta/Nb cation is displaced along the polar direction, indicating that isolated B-site off-centering does not correspond to an unstable polar displacement channel in the \textit{I4/mmm} reference phase. Individual displacements of Bi and Ca also lead to an increase in energy. In contrast, when the Bi$_{2}$O$_{2}$ layer is displaced collectively along the polar direction, the energy decreases at small finite $\delta$u$_{a}$ and forms a shallow minimum. This behavior demonstrates that the collective displacement of the Bi$_{2}$O$_{2}$ layer represents an energetically favorable soft displacement channel and serves as the primary structural pathway for the formation of polar structure in the Aurivillius layered compounds. It should be emphasized that the intrinsic instability of the polar $\Gamma_5^-$ mode is indispensable for the emergence of the ferroelectric phase. As shown in Table III, although the $\Gamma_5^-$X$_2^+$X$_3^-$ trilinear coupling provides the largest single contribution to the energy lowering, all of the biquadratic coupling terms among the $\Gamma_5^-$, X$_2^+$, and X$_3^-$ modes are positive. Their combined positive energy contribution even exceeds the stabilization arising from the trilinear coupling. Consequently, the $\Gamma_5^-$ distortion cannot be regarded as merely an induced secondary order parameter generated by coupling between the nonpolar distortions. Instead, its intrinsic soft-mode instability remains an essential prerequisite for the formation of the ferroelectric ground state. Based on the above results, we propose that the relative sliding between the Bi$_{2}$O$_{2}$ layer and the perovskite-like block represents a universal structural mechanism for the emergence of polarization in Aurivillius compounds. Since the Bi$_{2}$O$_{2}$ layer and the perovskite block constitute the fundamental building units of the Aurivillius structure, this mechanism is expected to be generally applicable throughout the Aurivillius family. Consequently, stable Aurivillius compounds typically exhibit ferroelectricity, including SrBi$_{2}$Ta$_{2}$O$_{9}$, SrBi$_{2}$Nb$_{2}$O$_{9}$, BaBi$_{2}$Ta$_{2}$O$_{9}$, Bi$_{4}$Ti$_{3}$O$_{12}$ etc., although their Curie temperatures and polarization strengths may differ substantially.

\subsection{Piezoelectric properties}
CBTO and CBNO both crystallize in a polar orthorhombic structure with the \textit{mm2} point group. For clarity and consistency with the polarization and stress directions used in our calculations, the following piezoelectric analysis is presented in the equivalent \textit{Cmc2$_{1}$} setting. Under this symmetry, the piezoelectric strain tensor \textit{d$_{ij}$} has five independent nonzero components: \textit{d$_{31}$}, \textit{d$_{32}$}, \textit{d$_{33}$}, \textit{d$_{24}$}, and \textit{d$_{15}$}. The piezoelectric constants are calculated using the direct finite-stress method. The polarization Pi is calculated under different applied stresses $\sigma_{j}$, and the corresponding \textit{d$_{ij}$} values are obtained from the linear dependence of P$_{i}$ on $\sigma_{j}$, as summarized in Table VI. The calculated piezoelectric tensors reveal a strongly anisotropic response in both CBTO and CBNO. For both of CBTO and CBNO, \textit{d$_{24}$} has the largest absolute value, indicating that a strong shear piezoelectric response driven by the shear stress $\sigma_{4}$. Under the \textit{Cmc2$_{1}$} setting, this coefficient corresponds to the coupling between the yz shear stress and the polarization along the y direction, reflecting a shear-induced polarization rotation within the yz plane. Within the yz plane, the symmetry of the parent phase permits four symmetry-equivalent polarization orientations. As a result, the energy landscape governing in-plane polarization rotation is relatively flat, enabling the polarization to rotate readily along low-energy pathways under the external field. This facile polarization rotation is responsible for the exceptionally large shear piezoelectric coefficient \textit{d$_{24}$}. The next relatively large tensor component is \textit{d$_{15}$}; however, it exhibits an unusually large negative value. This result indicates that shear deformation involving the out-of-plane direction induces a polarization response opposite to the positive crystallographic direction. This behavior is closely associated with the layered Aurivillius structure, where the insertion of Bi$_{2}$O$_{2}$ layers along the stacking direction interrupts the continuous BO$_{6}$ octahedral framework and suppresses long-range polar displacements. The longitudinal piezoelectric coefficient \textit{d$_{33}$} is the most important tensor component for practical piezoelectric applications, as it corresponds to the direct longitudinal deformation mode and is the most widely utilized operating mode in piezoelectric devices. Both CBTO and CBNO exhibit a relatively large positive \textit{d$_{33}$}, indicating an efficient longitudinal electromechanical coupling. By contrast, \textit{d$_{31}$} and \textit{d$_{32}$} possess much smaller absolute values, suggesting that the normal stresses $\sigma_{1}$ and $\sigma_{2}$ induce comparatively weaker polarization responses along the polar axis than $\sigma_{3}$. It should be noted that, compared with CBTO, the \textit{d$_{33}$} and \textit{d$_{32}$} coefficients of CBNO are significantly enhanced. This result indicates that Nb substitution strengthens the stress-polarization coupling along the polar axis, which is likely associated with the enhanced polar response of the perovskite block identified in the polarization decomposition. Overall, the calculated piezoelectric coefficients are consistent with experimental observations that CBNO ceramics exhibit superior piezoelectric performance compared with CBTO ceramics.

\begin{table}[t]
	\centering
	
	\begin{threeparttable}
		
		\caption{
			Piezoelectric constants of
			$\mathrm{CaBi_2Ta_2O_9}$ and $\mathrm{CaBi_2Nb_2O_9}$,
			with $d$ in units of $\mathrm{pC/N}$.
		}
		
		\label{tab:piezoelectric_constants}
		
		\small
		\renewcommand{\arraystretch}{1.20}
		\setlength{\tabcolsep}{3pt}
		
		\begin{tabular*}{\columnwidth}{
				@{\extracolsep{\fill}}
				l
				l
				c
				c
				c
				c
				c
				@{}
			}
			
			\specialrule{0.8pt}{0pt}{1.2pt}
			\specialrule{0.4pt}{0pt}{2pt}
			
			Materials
			&
			Method
			&
			$d_{31}$
			&
			$d_{32}$
			&
			$d_{33}$
			&
			$d_{24}$
			&
			$d_{15}$
			\\
			
			\midrule
			
			\multirow[c]{2}{*}{$\mathrm{CaBi_2Ta_2O_9}$}
			&
			PBE\tnote{a}
			&
			-4.9
			&
			4.5
			&
			19.1
			&
			37.7
			&
			-21.8
			\\
			
			&
			PBEsol\tnote{a}
			&
			6.83
			&
			3.19
			&
			15.72
			&
			32.01
			&
			-6.95
			\\
			
			\midrule
			
			\multirow[c]{2}{*}{$\mathrm{CaBi_2Nb_2O_9}$}
			&
			PBE\tnote{b}
			&
			-5.5
			&
			14.5
			&
			26.5
			&
			39.2
			&
			-20.6
			\\
			
			&
			PBEsol\tnote{b}
			&
			-1.15
			&
			6.45
			&
			17.83
			&
			64.13
			&
			-5.22
			\\
			
			\specialrule{0.4pt}{2pt}{1.2pt}
			\specialrule{0.8pt}{0pt}{0pt}
			
		\end{tabular*}
		
		\begin{tablenotes}[flushleft]
			\footnotesize
			\item[$^{a,b}$]
			Obtained by Tan et al.\ from DFT calculations\cite{16,44}.
		\end{tablenotes}
		
	\end{threeparttable}
\end{table}
Polycrystalline Aurivillius ceramics are of considerable importance for high-temperature piezoelectric applications because of their low fabrication cost, high fracture toughness, and suitability for large-scale manufacturing. The piezoelectric response of a polycrystalline ceramic can be regarded as the collective contribution from grains with different crystallographic orientations. After poling under a DC electric field, the spontaneous polarizations of individual grains become partially aligned, giving rise to a macroscopic net polarization and a measurable piezoelectric response. Consequently, the piezoelectric properties of polycrystals are closely related to the intrinsic piezoelectric coefficients of their constituent single crystals. Our previous studies have shown that, among the piezoelectric tensor components of Aurivillius ferroelectrics, the longitudinal coefficient \textit{d$_{33}$} and the in-plane shear coefficient \textit{d$_{24}$} make the dominant contributions to the piezoelectric performance of polycrystalline ceramics. 

By combining the piezoelectric constitutive relation with the polarization decomposition framework, the piezoelectric responses \textit{d$_{33}$} and \textit{d$_{24}$} can be quantitatively decomposed into contributions arising from Born effective charges and stress-induced atomic displacements, thereby revealing their microscopic atomic origins. Within the approximation that the stress-induced volume change is negligible, the ionic contribution to the piezoelectric constant can be written as:
\begin{equation}
	d_{ij}^{ion}=\frac{e}{\Omega }\sum\limits_{\kappa ,\alpha }{Z_{\kappa ,i\alpha }^{*}\frac{\partial \mu _{\kappa \alpha }^{\operatorname{int}}}{\partial {{\sigma }_{j}}}}\
	\label{eqn:S-criteria}
\end{equation}
where e is the elementary charge, $\Omega$ represents unit cell volume, $\kappa$ labels the atoms in the unit cell, $\alpha$ denotes the Cartesian direction. $Z_{\kappa ,i\alpha }^{*}$ is the Born effective charge tensor. $\mu _{\kappa \alpha }^{\operatorname{int}}$ is the atomic displacement induced by the applied stress, and $\sigma_{j}$ is the applied stress. This equation indicates that the piezoelectric response is governed by the product of the Born effective charges and the stress-induced internal atomic displacements. Therefore, large anomalous Born effective charges alone are insufficient to produce a strong piezoelectric response unless they are accompanied by substantial internal atomic relaxation under applied stress. The calculated results are summarized in Table VII. For \textit{d$_{33}$}, the Bi atoms in both CBTO and CBNO exhibit large Z$_{33}^*$ values together with significant $\sigma_{3}$ induced displacements. Consequently, they provide the dominant contribution to the longitudinal piezoelectric response along the polar axis. However, although Ta has a larger Born effective charge, its internal displacement response is much weaker, which substantially limits its contribution to \textit{d$_{33}$}. In contrast, Nb exhibits a much stronger internal displacement response under the same applied stress, providing a natural explanation for the larger \textit{d$_{33}$} of CBNO compared with CBTO. Furthermore, we find that the distribution of $\partial\mu_{3}$/$\partial\sigma_{3}$ closely follows that of the atomic displacements, suggesting that the longitudinal piezoelectric response is closely related to the polar $\Gamma_5^-$ mode. These results further demonstrate that the relative sliding between the Bi$_{2}$O$_{2}$ layer and the perovskite block plays an important role in enhancing the longitudinal piezoelectric response. In particular, the interfacial Bi and O3 atoms exhibit the largest stress-induced relative displacements associated with the layer sliding, making a substantial contribution \textit{d$_{33}$}. This finding indicates that the Bi$_{2}$O$_{2}$ layer is far more important for the piezoelectric response than has generally been recognized in previous studies, where the role of Bi was often underestimated because the Bi$_{2}$O$_{2}$ layer was treated as a rigid structural unit.

\begin{table}[t]
	\centering
	
	\caption{
		Born effective charges
		($Z_{33}^{*}$, $Z_{22}^{*}$) ($\lvert e\rvert$);
		$\partial \mu_{3}/\partial \sigma_{3}$ and
		$\partial \mu_{2}/\partial \sigma_{4}$
		($10^{-4}$~\AA/kBar).
	}
	
	\label{tab:born_effective_charges}
	
	\small
	\renewcommand{\arraystretch}{1.18}
	\setlength{\tabcolsep}{4pt}
	
	\resizebox{\columnwidth}{!}{%
		\begin{tabular}{@{}lccccc@{}}
			
			\specialrule{0.8pt}{0pt}{1.2pt}
			\specialrule{0.4pt}{0pt}{2pt}
			
			Materials
			&
			atom
			&
			$Z_{33}^{*}$
			&
			$\partial \mu_{3}/\partial \sigma_{3}$
			&
			$Z_{22}^{*}$
			&
			$\partial \mu_{2}/\partial \sigma_{4}$
			\\
			
			\midrule
			
			\multirow[c]{8}{*}{$\mathrm{CaBi_2Ta_2O_9}$}
			&
			$\mathrm{Ca}$
			&
			2.4927
			&
			3.2
			&
			2.5630
			&
			2.4
			\\
			
			&
			$\mathrm{Bi}$
			&
			4.7138
			&
			11.6
			&
			4.5987
			&
			12.2
			\\
			
			&
			$\mathrm{Ta}$
			&
			6.5148
			&
			1.5
			&
			7.5033
			&
			17.8
			\\
			
			&
			$\mathrm{O1}$
			&
			-3.2102
			&
			-9.4
			&
			-3.3351
			&
			-9.9
			\\
			
			&
			$\mathrm{O2}$
			&
			-3.1894
			&
			6.4
			&
			-3.9687
			&
			-10.3
			\\
			
			&
			$\mathrm{O3}$
			&
			-2.1390
			&
			-10.4
			&
			-2.1906
			&
			-4.6
			\\
			
			&
			$\mathrm{O4}$
			&
			-2.9588
			&
			3.2
			&
			-2.9027
			&
			-3.6
			\\
			
			&
			$\mathrm{O5}$
			&
			-1.9552
			&
			-9.2
			&
			-1.9730
			&
			-6.2
			\\
			
			\midrule
			
			\multirow[c]{8}{*}{$\mathrm{CaBi_2Nb_2O_9}$}
			&
			$\mathrm{Ca}$
			&
			2.6002
			&
			4.3
			&
			2.6002
			&
			2.2
			\\
			
			&
			$\mathrm{Bi}$
			&
			5.1141
			&
			14.6
			&
			5.1141
			&
			11.2
			\\
			
			&
			$\mathrm{Nb}$
			&
			8.5289
			&
			3.3
			&
			8.5289
			&
			18.1
			\\
			
			&
			$\mathrm{O1}$
			&
			-4.2035
			&
			-9.1
			&
			-4.2035
			&
			-9.9
			\\
			
			&
			$\mathrm{O2}$
			&
			-4.2035
			&
			3.4
			&
			-4.2035
			&
			-9.8
			\\
			
			&
			$\mathrm{O3}$
			&
			-2.3409
			&
			-11.8
			&
			-2.3409
			&
			-4.5
			\\
			
			&
			$\mathrm{O4}$
			&
			-3.1062
			&
			3.2
			&
			-3.1062
			&
			-3.0
			\\
			
			&
			$\mathrm{O5}$
			&
			-2.1782
			&
			-11.6
			&
			-2.1782
			&
			-6.3
			\\
			
			\specialrule{0.4pt}{2pt}{1.2pt}
			\specialrule{0.8pt}{0pt}{0pt}
			
		\end{tabular}%
	}
	
\end{table}
\begin{figure*}[btp]
	\includegraphics[width=\linewidth]{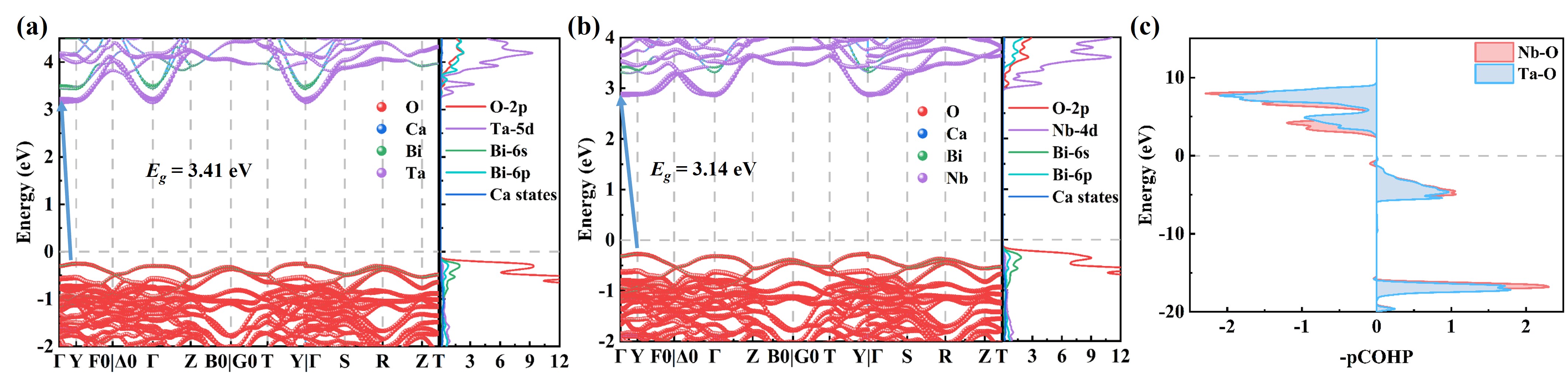}
	\caption{\label{fig5}Band structures (left) and projected density of states (right) of the orbitals of O/Ca/Bi/Ta atoms are calculated by the SCAN method for (a) CaBi$_{2}$Ta$_{2}$O$_{9}$ and (b) CaBi$_{2}$Nb$_{2}$O$_{9}$ in the \textit{A2$_1$am} phase. (c) Projected crystal orbital Hamilton population (-pCOHP) curves for the nearest-neighbor Ta-O and Nb-O bonds in the BO$_{6}$ octahedra of CaBi$_{2}$Ta$_{2}$O$_{9}$ and CaBi$_{2}$Nb$_{2}$O$_{9}$, respectively.}
\end{figure*} 
For \textit{d$_{24}$}, the present decomposition shows some difference from the full piezoelectric constants discussed above. Here, Z$_{22}^*$ and $\partial\mu_{2}$/$\partial\sigma_{4}$ are mainly used to identify the dominant response channel of \textit{d$_{24}$}. Since \textit{d$_{24}$} involves a shear-stress-induced transverse polarization change, its full value can also be affected by the off-diagonal components of the Born effective charge tensor and by internal displacements along other directions. The Ta/Nb atoms show strong transverse displacement responses under the shear stress $\sigma_{4}$, indicating that the BO$_{6}$ octahedra are key structural units for the shear piezoelectric response. This behavior is analogous to that of conventional perovskite ferroelectrics, where polarization rotation is dominated by the displacement of the B-site cations. In this framework, the relative sliding between the Bi$_{2}$O$_{2}$ layer and the perovskite block is no longer the dominant deformation mechanism. Instead, the atomic displacement response is characterized primarily by relative displacements between cations and anions. This behavior can be understood from the strongly anharmonic energy landscape created by the $\Gamma_5^-$X$_2^+$X$_3^-$ trilinear coupling, which effectively pins the spontaneous polarization. Instead, the limited in-plane polarization rotation is accommodated primarily by internal atomic displacements within the individual structural units, particularly the transverse displacements of the Nb/Ta and Bi atoms. Subsequently, the cooperative rearrangement of the oxygen sublattice further amplifies the polarization change along P$_{2}$ direction. Overall, Bi plays a crucial role in both the \textit{d$_{33}$} and \textit{d$_{24}$} piezoelectric responses and therefore cannot be neglected in understanding the piezoelectric mechanisms of CBTO and CBNO. The present analysis clarifies the microscopic origins of the piezoelectric response in these Aurivillius ferroelectrics and demonstrates that the superior piezoelectric performance of CBNO primarily originates from the stronger stress-induced displacement response of Nb compared with Ta.

\subsection{Origin of the intrinsic high resistivity}
A high electrical resistivity is an essential requirement for high-temperature piezoelectric materials because it prolongs the retention of stress-induced piezoelectric charges and prevents their rapid dissipation through thermally activated electrical conduction. Both CBTO and CBNO exhibit high electrical resistivities in experiments. However, at the same temperature, the resistivity of CBTO is typically one to two orders of magnitude higher than that of CBNO. To clarify the microscopic origin of the high resistivity in CBTO and CBNO, we compare their band structures calculated with different exchange-correlation functionals [Table SI]. We further analyze the orbital-projected density of states and band-edge effective masses at the SCAN level. The results show that both compounds are wide-band-gap insulators. The Fermi level lies within a well-defined band gap, and no partially occupied conducting bands appear near the band edges. Although PBE and PBEsol usually underestimate the band gap, the calculated band gaps remain about 2.9 eV for CBTO and 2.7-2.9 eV for CBNO. With the SCAN functional, the band gaps further increase to 3.41 eV and 3.14 eV, respectively. These results indicate that the insulating character is rather insensitive to the choice of functional. It is therefore a direct consequence of the intrinsic electronic structure.
From the viewpoint of charge transport, the intrinsic carrier concentration approximately follows\cite{45}:
\begin{equation} 
	{{\text{n}}_{i}}\propto \exp \left( -\frac{{{E}_{g}}}{2{{k}_{B}}T} \right)\
	\label{eqn:tri-CMT} 
\end{equation}
where E$_{g}$ is the band gap, k$_{B}$ is the Boltzmann constant, and \textit{T} is the temperature. Therefore, a band gap of about 3 eV strongly suppresses thermal excitation from the valence band to the conduction band at room temperature, leading to an exponentially small intrinsic carrier concentration. Compared with transport prefactors such as mobility or scattering time, the band gap plays a more fundamental role in determining the intrinsic carrier concentration and resistivity. Thus, in ideal intrinsic CBTO and CBNO, the high resistivity mainly originates from the extremely low intrinsic carrier concentration caused by the wide band gap.

The orbital-projected density of states further reveals the electronic origin of the band gap [Fig.~\ref{fig5}]. In both compounds, the valence-band maximum is mainly composed of O 2p states, with some hybridization from Bi 6s/6p states. The conduction-band minimum is dominated by empty Ta 5d or Nb 4d states, whereas the Ca-related states contribute little near the band edges. This band-edge character is consistent with the formal valence picture. In CBTO and CBNO, the B-site ions both correspond to the d0 electronic configuration. Their d states are therefore empty in the ground state and located above the Fermi level. The lowest interband excitation in the two compounds can be understood as a transition from the occupied O 2p valence band to the unoccupied B-site transition-metal d conduction band, namely O 2p → Ta 5d or O 2p →Nb 4d excitation. The large energy separation between the ligand O 2p states and the B-site metal d states is the key orbital mechanism for the wide-band-gap insulating behavior. The pCOHP analysis further support this view [Fig.~\ref{fig5}(c)]. The negative -pCOHP signals in the conduction band region indicate that the B-O interactions near the conduction-band edge have antibonding character. This result further confirms that the conduction band minimum should not be regarded as an isolated Ta/Nb d state, but rather as a B d-O 2p antibonding state derived from p-d hybridization. The difference between the band gaps of CBTO and CBNO can then be understood from the energy position of these B-site-derived antibonding conduction states. Since the valence-band maximum of both compounds is dominated by O 2p states, the larger band gap of CBTO mainly originates from the higher-lying Ta-derived conduction-band edge relative to the O 2p valence band. This can be rationalized by the more spatially extended nature of Ta 5d orbitals, which generally enhances Ta 5d-O 2p hybridization and increases the antibonding splitting compared with Nb 4d-O 2p hybridization. Consequently, the O 2p-Ta 5d charge-transfer separation is larger than the O 2p-Nb 4d separation, leading to the larger band gap of CBTO.
\begin{table}[t]
	\centering
	
	\caption{
		Effective hole masses at the $Y$ point and electron masses at the
		$\Gamma$ point for $\mathrm{CaBi_2Ta_2O_9}$ and
		$\mathrm{CaBi_2Nb_2O_9}$ calculated using the SCAN functional,
		in units of the electron rest mass $m_0$.
	}
	
	\label{tab:effective_masses}
	
	\small
	\renewcommand{\arraystretch}{1.25}
	\setlength{\tabcolsep}{4pt}
	
	\begin{tabular*}{\columnwidth}{
			@{\extracolsep{\fill}}
			l
			c
			c
			c
			@{}
		}
		
		\specialrule{0.8pt}{0pt}{1.2pt}
		\specialrule{0.4pt}{0pt}{2pt}
		
		Materials
		&
		Carrier types
		&
		Directions
		&
		Effective mass ($m_0$)
		\\
		
		\midrule
		
		\multirow[c]{6}{*}{$\mathrm{CaBi_2Ta_2O_9}$}
		&
		\multirow[c]{3}{*}{Electrons}
		&
		$\Gamma \rightarrow Y$
		&
		5.339
		\\
		
		&
		&
		$\Gamma \rightarrow F_{0}$
		&
		0.718
		\\
		
		&
		&
		$\Gamma \rightarrow Z$
		&
		0.544
		\\
		
		&
		\multirow[c]{3}{*}{Holes}
		&
		$Y \rightarrow \Gamma$
		&
		1.992
		\\
		
		&
		&
		$Y \rightarrow F_{0}$
		&
		0.927
		\\
		
		&
		&
		$Y \rightarrow S$
		&
		0.947
		\\
		
		\midrule
		
		\multirow[c]{6}{*}{$\mathrm{CaBi_2Nb_2O_9}$}
		&
		\multirow[c]{3}{*}{Electrons}
		&
		$\Gamma \rightarrow Y$
		&
		9.228
		\\
		
		&
		&
		$\Gamma \rightarrow F_{0}$
		&
		2.754
		\\
		
		&
		&
		$\Gamma \rightarrow Z$
		&
		0.828
		\\
		
		&
		\multirow[c]{3}{*}{Holes}
		&
		$Y \rightarrow \Gamma$
		&
		2.333
		\\
		
		&
		&
		$Y \rightarrow F_{0}$
		&
		4.590
		\\
		
		&
		&
		$Y \rightarrow S$
		&
		4.412
		\\
		
		\specialrule{0.4pt}{2pt}{1.2pt}
		\specialrule{0.8pt}{0pt}{0pt}
		
	\end{tabular*}
	
\end{table}

We also check the band-edge effective masses using a parabolic-band approximation, as the carrier effective mass is another important factor governing the electrical resistivity. Since electrical conduction in these wide-band-gap compounds is dominated by intrinsic carriers at elevated temperatures, the electron and hole effective masses along different directions are extracted from the band dispersion near the band extrema\cite{46,47}:
\begin{equation}
	\frac{1}{{{m}^{*}}}=\frac{1}{{{\hbar }^{2}}}\frac{{{\partial }^{2}}E(k)}{{{\partial }^{2}}k}\
	\label{eqn:multi_rabi}
\end{equation}
where E(k) is the band energy near the valence-band maximum (VBM) or conduction-band minimum (CBM). The effective mass is very sensitive to the band curvature, with flatter band giving rise to larger effective mass. The calculated electron effective mass of CBTO reaches 5.339 m$_{0}$ along the $\Gamma$→Y direction. In CBNO, the electron effective mass along the same direction further increases to 9.228 m$_{0}$. The hole effective masses of CBNO also reach 4.590 m$_{0}$ and 4.412 m$_{0}$ along the Y→F$_{0}$ and Y→S directions, respectively. These values indicate that the band edges in both compounds are weakly dispersive and that the carrier transport is strongly anisotropic. For layered bismuth oxides, this behavior can be related to the directional electronic coupling caused by the alternating stacking of Bi$_{2}$O$_{2}$ layers and perovskite-like slabs. It is also affected by the modulation of O 2p-B d orbital overlap by Ta/NbO$_{6}$ octahedral distortions. The weak orbital overlap makes the band edges relatively flat, which increases the effective masses and reduces the carrier mobility.

Therefore, the high resistivity of CBTO and CBNO has a clear intrinsic electronic-structure origin. The wide O 2p-Ta/Nb d charge-transfer gap strongly suppresses thermal excitation from the valence band to the conduction band and determines the main order of magnitude of the resistivity. The weak band-edge dispersion and large anisotropic effective masses further limit the mobility of the excited carriers. Thus, the high resistivity of these two compounds results from the combined effect of a wide band gap and low-mobility band-edge transport in d0 transition-metal bismuth-layered ferroelectric oxides. Therefore, the high resistivity of CBTO and CBNO can be attributed to their intrinsic electronic structures. Moreover, the comparison between CBTO and CBNO further supports this interpretation. Experimentally, CBTO ceramics usually exhibit a resistivity one to two orders of magnitude higher than that of CBNO ceramics. This behavior is consistent with the calculated larger band gap of CBTO, indicating that the Ta-induced widening of the O 2p-Ta 5d charge-transfer gap plays a dominant role in suppressing carrier excitation and enhancing the intrinsic insulating behavior.

\section{conclusion}
In summary, we combine group theoretic analysis and first-principles calculations to systematically investigate the ferroelectric phase-transition mechanism, the polarization, the piezoelectric response, and the intrinsic high resistivity of the Aurivillius layered perovskite ferroelectrics CBTO and CBNO. Our results show that the polar \textit{A2$_1$am} phase is not driven by a single polar soft mode. Instead, it results from the cooperative condensation of the polar $\Gamma_5^-$ mode, the X$_2^+$ oxygen octahedral rotation mode, and the X$_3^-$ oxygen octahedral tilting mode. The Landau free-energy analysis further reveals that the $\Gamma_5^-$X$_2^+$X$_3^-$ trilinear coupling plays a central role in stabilizing the ferroelectric phase, which the coupling provides an important energy-lowering channel and strongly reshapes the multidimensional potential-energy surface. This coupling therefore offers a microscopic explanation for the high-temperature stability of the ferroelectric state in these compounds. Berry phase calculations and Born effective charge analysis show that both CBTO and CBNO possess large spontaneous polarizations. The polarization arises from the combined effects of the interlayer sliding between the Bi$_{2}$O$_{2}$ layer and the perovskite-like block and the intralayer polar distortions within each structural unit. The piezoelectric tensors reveal a strongly anisotropic in both compounds. The ionic contribution plays an important role and mainly comes from the coupling between stress-induced internal atomic displacements and anomalous Born effective charges. The electronic-structure calculations further show that both compounds are wide-band-gap insulators. The large O 2p-Ta/Nb d charge-transfer gap suppresses intrinsic carrier excitation, while the weak band-edge dispersion and large anisotropic effective masses further limit carrier mobility. These two factors provide an intrinsic electronic structure origin for the high resistivity of CBTO and CBNO. These results establish a unified microscopic picture linking lattice distortions, polarization, piezoelectricity, and electronic insulation in CBTO and CBNO. They also provide theoretical guidance for designing layered ferroelectric materials with high Curie temperatures and robust insulating behavior.

~

\begin{acknowledgments}
This work was supported by the National Natural Science Foundation of China (Grant No.12404110).
The authors are also grateful for the useful discussion with Huazhang Zhang.
\end{acknowledgments}

\bibliography{APS-Bending.bib}
\end{document}